\documentstyle[12pt,aaspp4,flushrt]{article}
\begin{document}
\newcommand{\e}[1]{\eta_{#1}}
\newcommand{\divv}{\nabla\cdot{\mathbf{v}}}
\newcommand{\Ho}{H_\circ}
\newcommand{\hmpc}{h^{-1}\mbox{Mpc}}
\newcommand{\Osix}{\Omega_0^{0.6}}
\newcommand{\iras}{{\sl IRAS\/}}
\newcommand{\etal}{et al.}
\newcommand{\ltsima}{$\; \buildrel < \over \sim \;$}
\newcommand{\lsim}{\lower.5ex\hbox{\ltsima}}
\newcommand{\gtsima}{$\; \buildrel > \over \sim \;$}
\newcommand{\gsim}{\lower.5ex\hbox{\gtsima}}
\newcommand{\erf}{{\rm erf}}
\newcommand{\erfc}{{\rm erfc}}
\newcommand{\chka}{$\sqrt{}$}
\newcommand{\chk}{\raise.5ex\hbox{\chka}}

\title{Using Cluster Abundances and Peculiar Velocities to Test \\
the Gaussianity of the Cosmological Density Field}
\author{Weihsueh A. Chiu}
\affil{Joseph Henry Laboratories, Department of Physics, 
	Princeton, NJ 08544}
\author{Jeremiah P. Ostriker and Michael A. Strauss\footnote{Alfred
P. Sloan Foundation Fellow, and Cottrell Scholar of Research Corporation.}}
\affil{Princeton University Observatory, Princeton, NJ 08544}

\slugcomment{Printed \today: Revised 26 Aug 1997}

\begin{abstract}
Comparing the frequency of typical events with that of unusual events
allows one to test whether the cosmological density 
distribution function is consistent with the 
normally made assumption of gaussianity.  To this end, we
compare the consistency of the tail-inferred (from clusters) and measured 
values (from large-scale flows) of the rms level of mass fluctuations 
for two distribution functions: a Gaussian, and a
texture (positively-skewed) PDF.
We find that if we average the recent large-scale flow measurements, 
observations of the rms and the tail at the $10\hmpc$\ scale 
disfavor a texture PDF at $\sim 1.5\sigma$ in all cases.
If we take the most recent measurement of the rms,
that from  Willick \etal\ (1997b), the comparison 
disfavors textures for low
$\Omega_0= 0.3$, and disfavors Gaussian models if
$\Omega_0= 1$ (again at $\sim 1.5\sigma$). Predictions for evolution of
high temperature clusters can also be made for the models considered,
and, as is known (e.g., Henry 1997), 
strongly disfavor $\Omega_0 = 1$ Gaussian models, while
we find $\Omega_0 = 1$ marginally disfavored in  
texture models.  Taking the suite of
tests as a whole, and using all of the quoted data, it appears that
textures are strongly disfavored and only the low $\Omega_0$ Gaussian
models are consistent with all the data considered.  But given evidence for
the internal inconsistency of the observational data, had we only used
the recent Willick \etal\ results, the strength of our conclusion would
have been reduced to the $\sim 1\sigma$ level. 
\end{abstract}

\keywords{cosmology: large-scale structure of universe ---
galaxies: clusters and statistics ---
methods: analytical
}

\section{Introduction}

All current scenarios for the formation of structure in the Universe 
assume that cosmic structures grow from initially small seed
perturbations.  A fundamental assumption underlying a large class of these
theories is that the initial Fourier modes have random phases. 
This implies that the statistics of the initial density field
$\rho_i({\mathbf{x}})$ are fully specified by the correlation function
$\xi_i(r)=\langle\rho_i({\mathbf{x}})\rho_i({\mathbf{x}}+
{\mathbf{r}})\rangle_{\mathbf{x}}$, or equivalently its Fourier
transform, the power spectrum $P_i(k)$. 
For any given smoothing length, then, the one-point probability
distribution function (PDF) of the initial density field $\rho_i$ is
Gaussian.  A measurement of the frequency of rare events (clusters of
galaxies) coupled with a measurement of the level of rms fluctuations
can tell us directly whether or not the distribution of density
fluctuations satisfy the Gaussian hypothesis.

While theories based on the inflationary paradigm generally predict
Gaussian initial conditions, there are many alternatives which do not.
Global textures and other topological defects, for example, predict a
strongly non-Gaussian distribution for the initial perturbations
(e.g., \cite{gst91}; \cite{pst91}; \cite{gooding92}; \cite{pen94};
\cite{penetal97}).
Global textures arise when a global non-Abelian symmetry, such as
SU(2), is spontaneously and completely broken.  When these defects
collapse, energy gradients in the Higgs field perturb the metric and
induce gravitational fluctuations in the matter fields.  These texture
knots, then, act as ``causal seeds'' of structure formation.  We take
the texture model as our standard non-Gaussian model, since it is well
studied and still viable as a theory of structure formation.  
In general, texture
and Gaussian models give similar predictions for the 
\emph{shape} of the power spectrum and the level of CMB fluctuations
(\cite{pen94}; \cite{pen95}).  In fact, it is often noted
(e.g., \cite{pst91}) that texture and Gaussian CDM models are
\emph{most} different in their PDFs.  If it could be demonstrated that
the initial density field were Gaussian on some suitable scale, the texture model would
be directly falsified.

The standard method for testing Gaussianity relies on  
measuring 
the $N$-th order reduced moments, or cumulants $\lambda_N$ of the
present-day PDF, such as skewness
or kurtosis, from redshift or peculiar velocity surveys (cf., Strauss
\& Willick 1995 for a review). 
The variance is $\sigma^2 \equiv \lambda_2\equiv
\langle\delta^2\rangle$, the skewness is
$\lambda_3\equiv\langle\delta^3\rangle$, and the kurtosis is
$\lambda_4\equiv\langle\delta^4\rangle - 3\sigma^4$, where $\delta$ is
the density contrast $\rho/\langle\rho\rangle - 1$.  For a Gaussian
random field, all the cumulants of order $N>2$ are equal to zero.  The
growth of Gaussian initial perturbations under gravitational
instability naturally produces non-Gaussianity in the distribution
function of the density field.  In particular, the cumulants follow a
scaling relation $\lambda_N = S_N \lambda_2^{N-1}$, where the $S_N$
are independent of scale in the mildly non-linear regime for power-law
power spectra (Fry 1984ab, Bernardeau 1992), and can be calculated
from $(N-1)$-th order perturbation theory (Bernardeau 1994).
Significant departures from scale-invariance are taken as signs of
non-Gaussianity.  Unfortunately, the higher-order moments are very
sensitive to finite-volume effects and non-Gaussian noise in the data
(c.f. the discussion in \cite{ks97} and references therein).

Since many non-Gaussian distribution functions have long tails, it is
better to \emph{directly} compare the rms and the tail of the PDF,
rather than relying on the moments, which are integrals over the PDF.
Given a distribution function, the level of tail fluctuations maps
directly to a level of rms fluctuations; the tail-inferred rms, then,
must be consistent with the directly measured rms.  Furthermore, one
should look for measures which are, as closely as possible, indicators
of the initial, linear density field.

In this paper, we use the observed temperature distribution of  
clusters of galaxies as a measure of the tail of the distribution 
of initial density fluctuations, and galaxy peculiar velocities and 
redshift-space distortions as measures of the rms.  
Clusters of galaxies are the most massive
virialized objects in the universe, with densities a hundred or more
times that of the mean density of the universe, containing only a few
percent of all galaxies.  They are thus 
extreme excursions from the mean, and hence are measures of the
high potential tail of the distribution of fluctuations.
We compare this with the rms level of 
peculiar velocity fluctuations.  The rms is not
as sensitive to non-linear growth as are higher order statistics
(Dekel, private communication).
Both cluster and peculiar velocity observations 
are at roughly the same spatial scale, $\sim 10\hmpc$.  
Furthermore, they both measure gravitational potential fluctuations,
and therefore have a similar dependence on $\Omega_0$, which
can be approximated by defining
\begin{equation}
\e{R}\equiv\sigma_R\Osix,
\end{equation}
where $\sigma_R$ is the 
rms mass fluctuation on \emph{tophat} filter scale $R.$  
The remaining $\Omega_0$ dependence is $\leq 5$\% for the 
peculiar velocity and redshift distortion measurements, 
and $\leq 25$\% in the range $0.3 \leq\Omega_0\leq 1.0$ 
for the cluster measurements.  This is the quantity we 
will attempt to determine throughout our analysis, comparing estimates
of it found by various methods. 

In \S 2, we examine various recent derivations of $\e{10}$
from the Mark III peculiar velocity catalog (Willick \etal\ 1995,
1996, 1997a), and from \iras\ galaxy redshift space distortions.   
In \S 3, we infer $\e{10}$ from the redshift-zero
X-ray cluster abundances, 
assuming a Gaussian and texture PDF, using a modified 
Press-Schechter approach.  
We calculate the $\e{10}$ separately for $\Omega_0=1.0$,
cosmologically flat ($\Lambda$-dominated) $\Omega_0=0.3$, and open
$\Omega_0=0.3$.   In \S 4, we discuss the consistency of the rms and
the tail-inferred measures of $\eta$ by performing a likelihood
analysis of the Gaussian and texture hypotheses.  
We also discuss further constraints from 
observations of clusters at moderate redshift.  
We summarize in \S 5.

\section{RMS Mass Fluctuations from Peculiar Velocities and Redshift
Distortions }

	It has been noted for more than a decade that the 
rms optical galaxy counts in $8\hmpc$\ spheres is 
$\left\langle(\delta N/N)_8^2\right\rangle^{\onehalf}\equiv \sigma_{8}^{\rm Optical} \approx 1$ 
(\cite{DP83}).  
But given that galaxy formation is
still not very well understood, it has become customary to relate 
this measured variance to the more fundamental rms fluctuations in the 
matter density $\sigma_8$ by a bias parameter $b$.  For optical
galaxies, $b^{\rm Optical}$ is defined by $\sigma_8^{\rm Optical} 
\equiv b^{\rm Optical}\sigma_8$.  The bias,
however, is clearly dependent on the galaxy type
(for example, as we will see below $\sigma_8^{\sl IRAS} = 0.69\pm
0.04$, \cite{Fisher94a}), and 
may be dependent on scale.  We wish to estimate the rms value of
fluctuations independent of bias.  To do this, we look at recent
analyses of peculiar velocity and redshift survey data. 

\subsection{Recent Measurements of RMS Fluctuations} 

	The peculiar velocity ${\mathbf{v}}$ 
is the deviation of the proper velocity 
of a galaxy from the local Hubble flow.  Assuming that this
measured velocity follows from gravitational potential fluctuations,
we obtain from linear perturbation theory (\cite{PPC}):
\begin{equation}
	\divv = -\Ho f(\Omega_0) \delta \approx -\Ho \Osix \delta.
	\label{eq-divv}
\end{equation}
Here $\delta$ is the mass density contrast 
$\delta({\mathbf{x}}) \equiv\rho_{\rm lin}({\mathbf{x}})/\bar\rho -
1$, and the approximation is good to 5\%.
Now consider filtering the velocity and density fields 
with a tophat of radius $R$,
yielding the filtered
fields ${\mathbf v}_R$ and $\delta_R$.
Then $\divv_R$, the divergence of the filtered velocity 
field, is proportional to 
the filtered mass density field $\delta_R$.
Taking the rms values of $\divv_R$ and $\delta_R$ gives:
\begin{equation}
	\Ho^{-1}\langle(\divv_R)^2\rangle^{\onehalf}
	\approx \Osix\langle\delta_R^2\rangle^{\onehalf}
	\equiv \Osix\sigma_R \equiv \e{R}.
\label{eq-divv-smoothed}
\end{equation}
The peculiar velocity field thus provides a direct measurement of
mass fluctuations, appearing with the cosmological density parameter
$\Omega_0$.  
Note that in linear theory, where individual Fourier components evolve
independently, it makes no difference in which order one filters and 
takes the divergence of the velocity field.

	Seljak \& Bertschinger (1993) made an 
early attempt to measure $\eta_R$ directly 
using the POTENT method (Bertschinger \etal\ 1990) 
to reconstruct the mass density and peculiar 
velocity fields from the Mark II catalog (Burstein 1989). 
They derived a value of $\e{8}=1.3\pm 0.3$.  More recently,
several workers have used the Mark III catalog of peculiar velocities
(Willick \etal\ 1995, 1996, 1997a)
to derive values of $\eta$.  

Zaroubi \etal\ (1997) perform a
likelihood analysis of the peculiar velocities given families of 
CDM-type models.  Because we are making a comparison only 
with clusters of galaxies, which are measures of fluctuations on 
scales of roughly 10 $\hmpc$, we only consider their
COBE-independent results.  
Zaroubi \etal\ (1997) work with the measured peculiar velocities of
individual galaxies, without any smoothing, so 
it is unclear to which scale they are most 
sensitive.  We simply 
integrate their power spectra with $10\hmpc$ tophat smoothing
to obtain the rms fluctuation on that scale.  Their
``maximum likelihood'' model has $\e{10}=0.70$, and the
models on their ``65\%-confidence'' level contour span the range
$\e{10}=0.61-0.78$.  We therefore take their measurement to be
$\e{10}= 0.70\pm 0.09$.  

Kolatt \& Dekel (1997) use the POTENT technique (Dekel \etal\ 1990) to
determine the quantity $\divv_R$ from the 
Mark III peculiar velocities,
to compute the mass power spectrum directly, 
subtracting out a model for the noise power spectrum derived
from mock catalogs.  Since
POTENT applies a Gaussian filter of filter length 12$\hmpc$, equivalent
to a $\sim 23\hmpc$ tophat filter (see Appendix), we integrate the
functional form they give for the observed power spectrum, taking into
account the uncertainties in the parameterization, to derive a 
value of $\e{23}=0.32\pm 0.11$.

  If we assume linear biasing, $\delta_{\rm galaxies}= b \delta_{\rm dark\
matter}$, comparison of peculiar velocity and galaxy density field
data via equation~(\ref{eq-divv}) allows one to determine the
quantity $\beta\equiv\Osix/b$,
the proportionality constant between
the peculiar velocity and galaxy density fields for a
particular set of galaxies. 
Note that if bias is independent of scale, $\beta$ should
be as well. 
Following equation~(\ref{eq-divv-smoothed}) and the definition of $b$,
we find that for a comparison with \iras\ galaxies: 
\begin{equation}
\e{R} = \beta_{\sl IRAS}\sigma_{R,\sl IRAS}.
\end{equation}  
For $\sigma_{R,\sl IRAS}$, we use the power law
analytic approximation of Fisher \etal\ (1994a):
\begin{equation}
\sigma_R = \left[\frac{72(r_\circ/R)^\gamma}
		{2^\gamma(3-\gamma)(4-\gamma)(6-\gamma)}
		\right]^{\onehalf}; \qquad
	r_\circ = 3.76^{+0.20}_{-0.23} \hmpc; 
		\qquad \gamma = 1.66^{+0.12}_{-0.09}.
\label{eq-fisher}
\end{equation}

	Several groups have determined $\beta$ from
comparisons of {\sl IRAS} density and peculiar velocity fields.
Willick \etal\ (1997b) perform a likelihood analysis of
Tully-Fisher observables of the Mark III data given a prior velocity
model, 
and
find $\beta^{\sl IRAS}=0.49\pm 0.07$.  Their
analysis uses an effective Gaussian smoothing of 
$\sim 4\hmpc$\, and so corresponds 
roughly to an 8$\hmpc$\ tophat.  Using $\sigma_8^{\sl IRAS}=0.69\pm
0.04$, we obtain $\e{8}=0.34\pm 0.05$.  In another approach, 
Sigad \etal\ (1997) compare the POTENT-derived 
peculiar velocity divergence field to the \iras\ galaxy density 
field (with effective tophat
smoothing of $\sim 23\hmpc$ due to the POTENT method) 
and derive $\beta=0.89\pm 0.12$.  Multiplying by
$\sigma_{23}^{\sl IRAS} = 0.29 \pm 0.02$ leads to $\e{23}=0.26\pm
0.04$.  

	Finally, as first pointed out by Kaiser (1987), one can use
the anisotropy of the \emph{redshift}-space correlation function
of a redshift survey to infer the value of $\beta$.  
Fisher \etal\ (1994b) measured the anisotropy of the {\sl IRAS} 1.2Jy
sample at an effective scale of $10-15\hmpc$,
obtaining $\beta=0.45^{+0.27}_{-0.18}$.  Using
the same procedure as above to convert $\beta$ to $\eta$, we obtain
$\eta_{12.5}=0.21\pm 0.11$.  
Cole \etal\ (1994, 1995) 
use instead the power spectrum anisotropy, at an effective scale of 
$\sim 35\hmpc$, to obtain $\beta=0.52\pm 0.13$, 
resulting in $\eta_{35}=0.11\pm 0.03$.  Yet another
method involves decomposing the redshift-space distribution into
spherical harmonics.  Fisher, Scharf, \& Lahav (1994) 
(c.f. \cite{HeavensTaylor95}) show that the inferred value of $\beta$
depends sensitively on the assumed power spectrum.  
Scaling their results for 
each of the power spectra
they use, we find that the 
results for $\eta$ roughly converge
at $30\hmpc$, with a 
value $\eta_{30}=0.20\pm 0.03$.  

\begin{figure}
\plotone{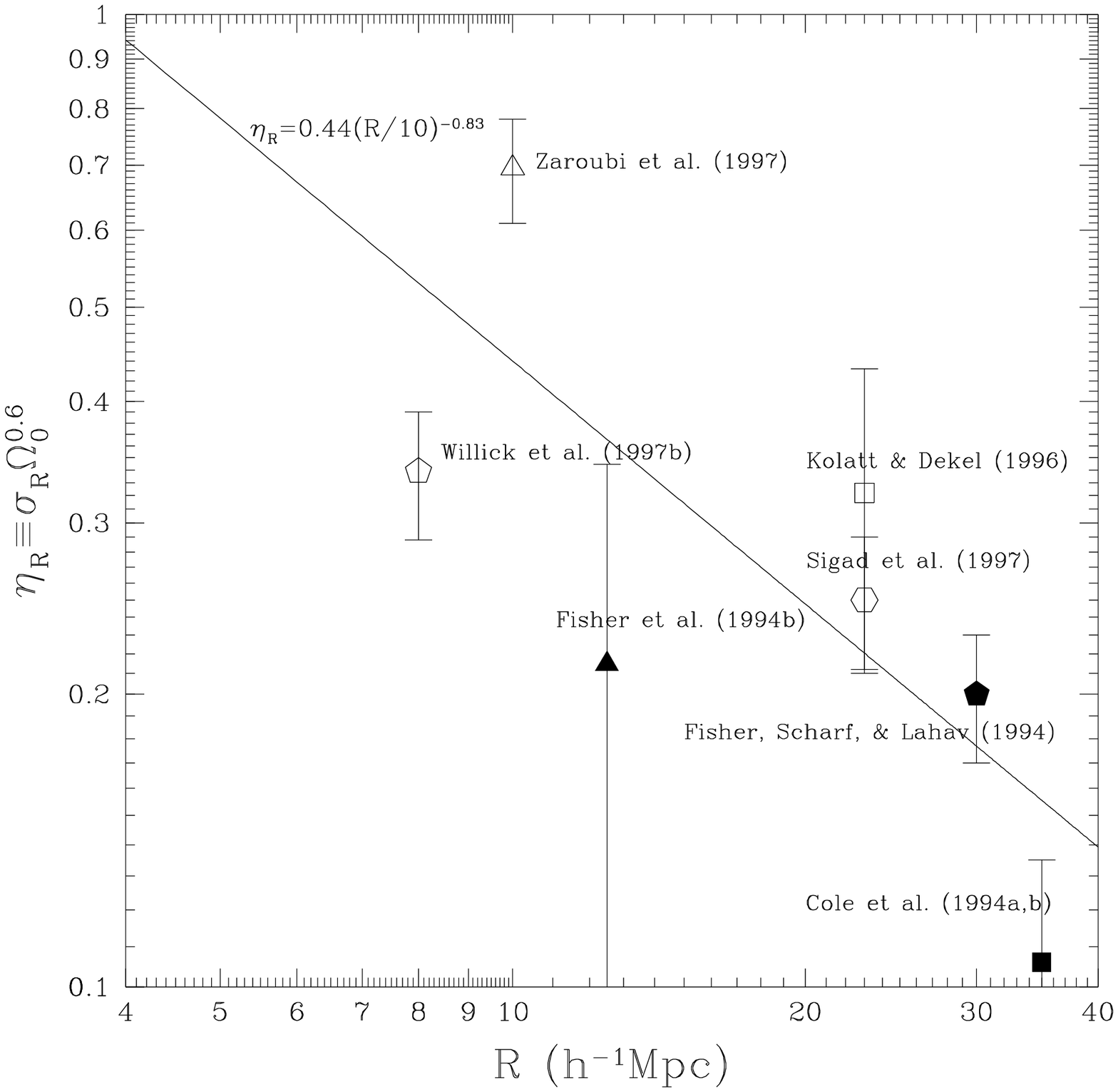}
\figcaption[nongaus_fig1.eps]{
RMS mass fluctuations as inferred from peculiar velocity
and redshift distortion analyses, as described in \S 2, and listed
in Table~\ref{tab:rms-eta}.
The solid line is $\e{R} = 0.44(R/10.0\hmpc)^{-0.83}$, where
$\alpha$ is that for \iras\ galaxies, and $\e{10}$ is 
taken from the maximum likelihood average in Table~\ref{tab:rms-eta}.
\label{fig:nongaus_etarms}
}
\end{figure}

	Table~\ref{tab:rms-eta} summarizes the values of $\e{R}$
cited above, all using rms fluctuations, while 
Figure~\ref{fig:nongaus_etarms} plots all the
recent measurements, 
with their respective errors, at their effective tophat filter
scales. Most
current theories of structure formation predict that $\sigma_R$ is
approximately a power law in the range 
$8 < R < 35\hmpc$.  We thus 
approximate $\e{R}$ as a power law $\e{R}\propto R^{-\alpha}$, to
allow us to extrapolate the results of Table~\ref{tab:rms-eta} and 
Figure~\ref{fig:nongaus_etarms}
to a common scale.  
In linear theory, 
a standard CDM model with $n=1$ and $\Omega_0=1$ has a logarithmic
slope at $10\hmpc$ of $\alpha\approx 1.0$, while a tilted
$n=0.7$, low density $\Omega_0=0.3$ CDM model has 
$\alpha\approx 0.6$.  An
$\Omega_0=1$ texture model (using the power spectra of \cite{pen94} 
and \cite{pen95}) gives $\alpha\approx 1.2$ at $10\hmpc$.
If galaxy bias does not change appreciably in this 
range, then $\alpha$ is the same as that for galaxies.  
Equation~(\ref{eq-fisher}) implies
an exponent $\alpha=0.83\pm 0.12$ for galaxy fluctuations.
If the true value of $\alpha$ for mass fluctuations 
is much less (greater) than 0.83, 
then bias must decrease (increase) with increasing scale.  
When extrapolating the above results to different scales, then, 
we take three different power laws to reflect this uncertainty
in the slope of the fluctuation spectrum, $\e{R}\propto R^{-\alpha}$: 
$\alpha = 0.83,\ 0.60$, and 1.10.  
Table~\ref{tab:rms-eta} contains extrapolations of 
$\e{R}$ to $10\hmpc$ for each of
these assumed values for $\alpha$.
It is striking that the different 
methods give answers which differ from one another by
significantly more than the quoted errors --- a sure sign that
systematic errors of unknown origin are important.  We now present a
method for combining these measurements. 

\subsection{An Error Model for the Different Measurements of $\eta$}

	Let us assume that each of the seven measurements 
(labeled $i=1\ldots 7$) listed 
in Table~\ref{tab:rms-eta} has an unknown systematic error $\mu_i$ from the 
true value of $\eta$.
That is, each one \emph{actually} measures $\eta+\mu_i$, rather
than $\eta$, where the stated statistical error $\epsilon_i$ is the
error in the measurement of $\eta+\mu_i$.  We assume that the
systematic error is drawn 
from a Gaussian distribution with zero mean and unknown 
variance $\theta^2$, and that the statistical errors $\epsilon_i$ are
also Gaussian distributed. 
The probability of a measurement $x_i$, given unknown
$\eta$ and $\theta$, is given by
\begin{eqnarray}
	{\cal P}(x_i|\eta,\theta) & \equiv & 
		\int d\mu_i {\cal P}(x_i|\mu_i,\eta) 
		{\cal P}(\mu_i|\theta)\\
	& = & \frac{1}{\sqrt{2\pi(\theta^2+\epsilon_i^2)}}
		\exp\left[-\frac{(x_i-\eta)^2}{2(\theta^2+\epsilon_i^2)}
		\right].
\end{eqnarray}
The systematic error simply adds an error $\theta$
in quadrature to each measurement error $\epsilon_i$.

\begin{figure}
\plotone{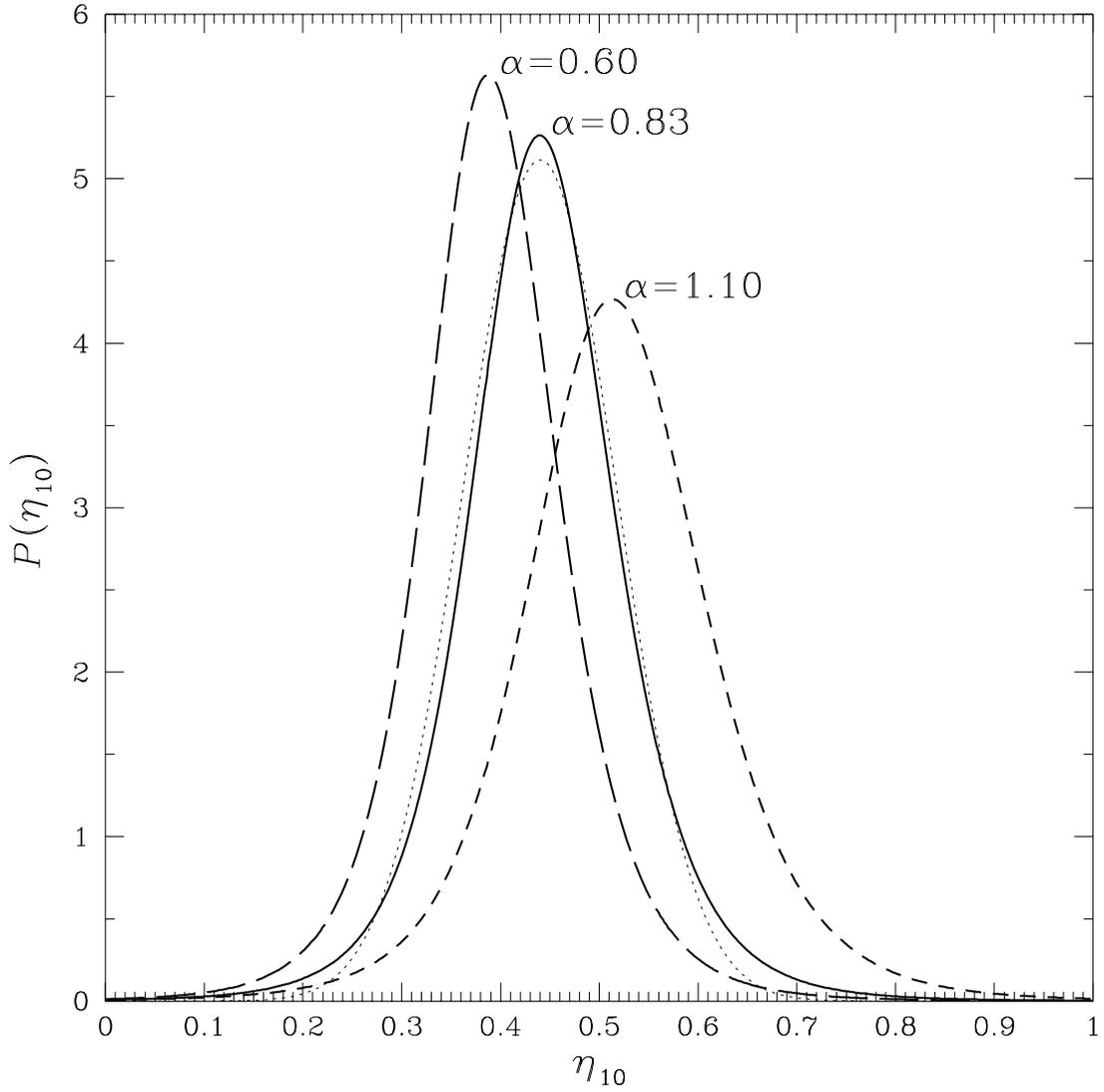}
\figcaption[nongaus_fig2.eps]{
Probability density for the error analysis in \S 2 of the
rms measurements of $\e{10}$, for power
law indices $\alpha=$0.60, 0.83, and 1.10.  
The thin dotted line is a Gaussian 
distribution with the same mean and 68\%
confidence interval ($1\sigma$ error) 
as the probability density for $\alpha=0.83$.  
\label{fig:nongaus_rmslike}
}
\end{figure}

If we define the likelihood function of the unknowns $\eta$ and
$\theta$ in the usual way, we obtain
\begin{eqnarray}
	\ln{\cal L}(\eta,\theta) & \equiv & \ln\left[\prod_i 
		{\cal P}(x_i|\eta,\theta)\right] \\
	& = & - \frac{1}{2} \sum_i\left[
		\frac{(x_i-\eta)^2}{\theta^2+\epsilon_i^2} +
		\ln\left[2\pi(\theta^2 + \epsilon_i^2)\right]\right].
		\label{eq-likerms}
\end{eqnarray}
The maximum likelihood value for $\eta$ is obtained by setting 
the derivative of equation~(\ref{eq-likerms}) to zero:
\begin{equation}
	\hat{\eta} = \left(\sum_i \frac{x_i}{\theta^2+\epsilon_i^2}\right) 
		\times \left(\sum_i 
			\frac{1}{\theta^2+\epsilon_i^2}\right)^{-1}.
		\label{eq-etamaxlike}
\end{equation}
Note that equation~(\ref{eq-etamaxlike})
gives the traditional weighted average for $\theta \ll \epsilon_i$,
and an unweighted average for $\theta \gg \epsilon_i$.  
Since we are primarily concerned with $\eta$, we numerically
integrate the likelihood ${\cal L}(\eta,\theta)$ over 
$\theta$ to obtain the marginal likelihood over $\eta$:
\begin{equation}
	{\cal L}_{\rm rms}(\eta) =
		\int_0^\infty {\cal L}(\eta,\theta) d\theta.
		\label{eq-prob_eta}
\end{equation}
In Figure~\ref{fig:nongaus_rmslike}, we show the likelihood
$ {\cal L}_{\rm rms}(\eta)$, each normalized to 
$\int d\eta\, {\cal L} = 1$, in the case of $\alpha=$0.60, 0.83, 
and 1.10, along with 
a Gaussian distribution of the same 68\% confidence level 
in the case of $\alpha=0.83$.  Note the significant 
tail to the distribution relative to a Gaussian, reflecting
the fact that the measurements are not consistent their 
individual normal errors.  
The values of $\e{10}$ corresponding to the 
maximum of ${\cal L}_{\rm rms}(\eta)$, equation~(\ref{eq-prob_eta}),
along with $1\sigma$ (68\% confidence) errors, are given in 
the last row of Table~\ref{tab:rms-eta}, for various $\alpha$.  
For $\alpha = 0.83$, the rms
measurements of the velocity field give 
$\eta_{10} = 0.44\pm 0.08\ (1\sigma)$. 

	This model is only a rough attempt at quantifying the degree
of systematic errors present in these measurement.  
The assumptions that each 
systematic error $\mu_i$ is independent and normally distributed 
are certainly not completely correct.  All of the analyses
use the {\sl IRAS} galaxy redshift catalog, and many of them use
the same Mark III catalog, and hence their results must be somewhat
correlated.  Furthermore, systematic errors are rarely normally
distributed.  Perhaps one of the measurements is correct, and all the
others are erroneous.  In comparing rms and tail-inferred values of $\eta$, 
we therefore also use 
the Willick \etal\ (1997b) result alone, which is the most recent
result which explicitly filters
at a scale close to the $\sim 10\hmpc$ scale from which clusters
form. 

\section{Tail Fluctuations: Abundances of X-ray Clusters}

	Galaxy statistics and peculiar velocities measure
the rms fluctuations of the density field --- i.e., fluctuations
within $\sim 1\sigma$ from the mean.  Most probability distributions look
similar to Gaussians at this fluctuation level --- they are peaked
and convex near the mean.  The more dramatic differences between
Gaussian and non-Gaussian distributions come in their tails.  Rich
clusters of galaxies, the most massive virialized objects in the
universe, offer the best measure of the tail of the probability
distribution of the density field on scales of $\sim 10\hmpc$.  
Approximately 5\% of the $L_*$ galaxies
are within one Abell radius of a rich cluster (\cite{Bahcall96}), 
and about 10\% of the baryons
reside in clusters (taking a BBN value of $\Omega_b = 0.0125h^{-2}$ 
from \cite{walker91}).  Thus in the Gaussian scenario, they constitute 
$\gtrsim 1.5\sigma$ perturbations.  
They are well-observed by a number of techniques; in particular, 
their abundance as a function of X-ray temperature is well-measured
(Henry \& Arnaud 1991). 
Thus they provide an accurate estimate of the integral over the 
tail of the distribution function at the $\gtrsim 1.5\sigma$ level.  

The temperature function of rich clusters can be predicted from the
statistics of peaks in a density field; it depends on the
rms level of fluctuations $\sigma_R$ on the scale on which clusters
form, as well as $\Omega_0$ through the relation between comoving
radius and mass.  Comparing these predictions with observations
constrains the quantity
$\sigma_R\Omega_0^{\nu},\nu = 0.4 \sim 0.6$ 
(cf., White, Efstathiou, \& Frenk 1993; Eke, Cole, \& Frenk 1996; Pen
1997, and references therein). 
The mass of a rich cluster within the $1.5\hmpc$ Abell radius
is $10^{14}-2\times 10^{15}h^{-1}M_{\sun}$, 
corresponding to a region with
initial comoving radius $R=[3M/(4\pi\bar\rho_\circ)]^{\onethird}=(4.4-12)
\Omega_0^{-\onethird}\hmpc$.  
Numerical simulations have confirmed that the cosmic abundance of these
objects is determined by the power spectrum at 
$k^{-1} = 5-20\hmpc$, so that their frequency provides a
direct measure of $\sigma_{R}$ in this range (e.g., Bahcall \& Cen
1992; \cite{Evrard96}; Cen 1997, in preparation).
Hence the cluster abundances constrain $\e{10}$ with 
only a weak dependence on $\Omega_0$.
In \S 3.1 we derive the relation between the distribution of
clusters as a function of temperature and the assumed density PDF.
This requires a relation between the observed temperature of a cluster
and the initial comoving radius from which it formed; this is given in
\S 3.2.  This relation depends on the formation epoch of the cluster,
as derived in \S 3.3.  We compare these results with the observed
temperature distribution of clusters in \S 3.4 to derive the quantity
$\e{10}$. 

\subsection{Press-Schechter and the Temperature Function}

We use the Press-Schechter (1974) \emph{ansatz} (hereafter PS) to 
derive the number density of clusters.  Many have done this in the
Gaussian case (e.g., White \etal\ 1993), and here 
we make a straight-forward generalization to non-Gaussian distributions.
In PS, an object of mass $\geq M$ forms when the
linearly-extrapolated density contrast filtered on a mass scale $M$,
corresponding to a tophat comoving length scale $R$,  
exceeds the threshold 
$\delta_c \approx 1.69$.  The exact value of $\delta_c$ depends on 
$\Omega_0$ and $\Lambda$, and can be derived assuming uniform spherical
collapse (see \cite{ECF}), 
but varies by only a few percent 
for $\Omega_0 > 0.2$.
The differential abundance from PS, given here as the comoving 
number density at the present of objects with initial comoving
radius in the interval $(R,R+dR)$, can be written as 
\begin{equation}
	 n_RdR 
		= \frac{3}{4\pi R^3}\left |\frac{d}{dR}
			\int_{\delta_c/\sigma_R}^\infty
			f\, P(y)dy
			\right | dR,
\end{equation}
where we assume that the mass distribution function $P(y)$ only depends on
$y \equiv \delta/\sigma_R$. 
This is true in Gaussian theories; in
the texture model, the seeding of density perturbations is
scale-invariant, and thus the distribution function must also be
scale-invariant. 
The prefactor $f\equiv[\int_0^\infty P(y)dy]^{-1}$ normalizes $P$ so
that all the mass in the universe is accounted for.  
Assuming, as before, a power law dependence 
$\sigma_R \propto R^{-\alpha}$ gives 
\begin{equation}
	n_R = \frac{3}{4\pi R^3} \frac{\alpha y}{R} f\, 
			P(y),
		\quad y \equiv \frac{\delta_c}{\sigma_R}.
	\label{eq-nr}
\end{equation}
The initial comoving radius is of course not an observable, so we will
convert to the abundance as a function of present-day temperature:
\begin{equation}
	n_T\frac{dT}{dR} = \frac{3}{4\pi R^3} \frac{\alpha y}{R} 
			f\, P(y).
	\label{eq-ntps}
\end{equation}
The observed X-ray temperature function gives the $n_T$ of 
equation~(\ref{eq-ntps}) as a function of temperature.  
In order to find $y$ (and
hence $\sigma$ and $\eta$), we need the appropriate PDF for each
model, and the relationship between
the initial comoving radius and the temperature of
the clusters (next section).

\begin{figure}
\plotone{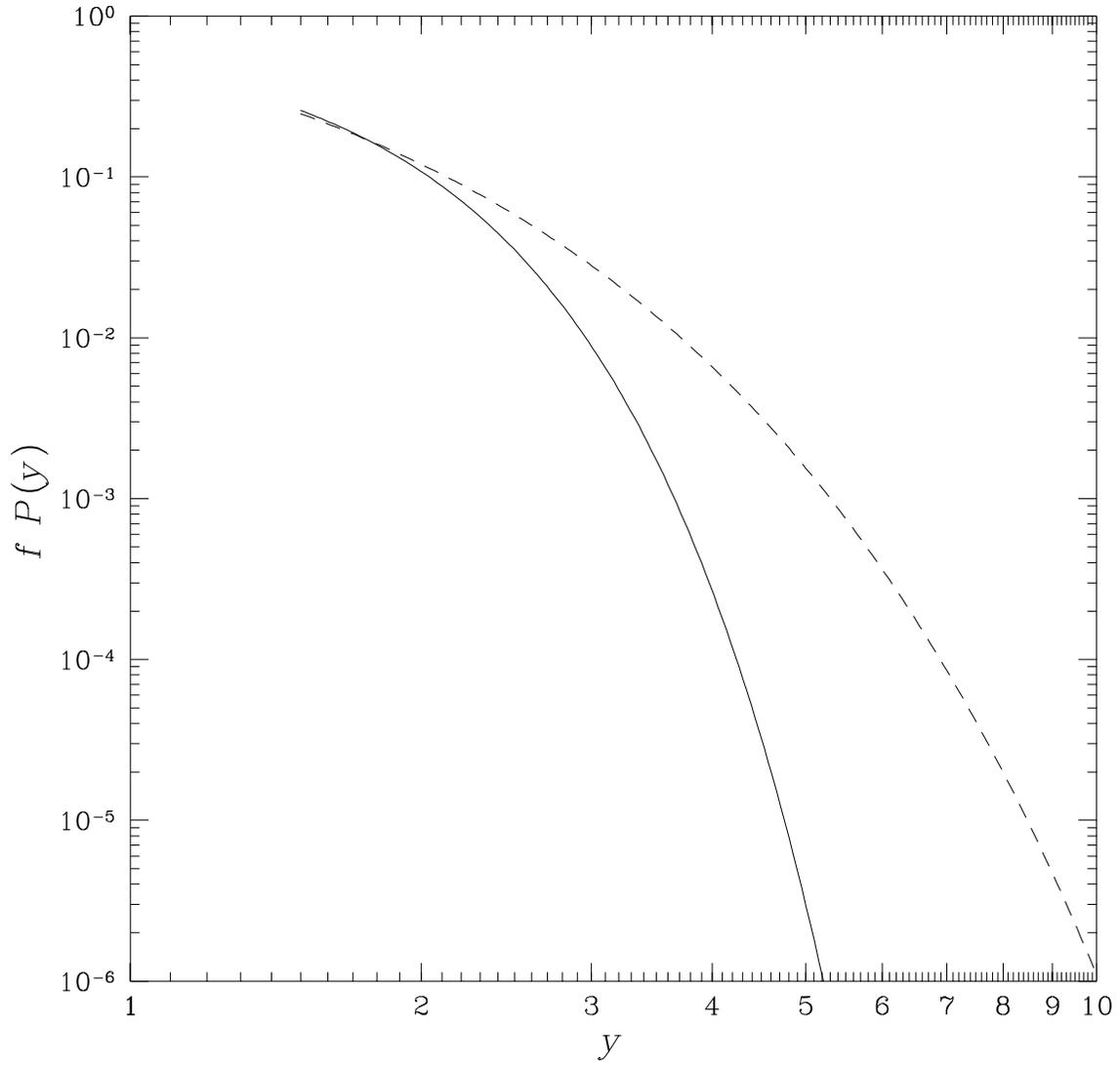}
\figcaption[nongaus_fig3.eps]{
Tails of the effective distribution functions $f\, P(y)$, for a
Gaussian (solid) and texture (dashed) PDF.
\label{fig:nongaus_pdfs}
}
\end{figure}

In Gaussian theories, the distribution function is 
\begin{equation}
	f\, P(y)=\frac{2}{\sqrt{2\pi}}\exp(-y^2/2),
	\label{eq-gauspdf}
\end{equation}
recalling that the factor $f\equiv[\int_0^\infty P(y)dy]^{-1} =2$.
For textures, we take the distribution function from the 
numerical simulations of Park, Spergel, \& Turok (1991), 
which are the only published simulations which are at 
scales appropriate for clusters of galaxies.
We find that the high-$y$ tail of their distribution gives: 
\begin{equation}
	f\, P(y) = 
	2.2\exp(-1.45y), \quad y \geq 1.5.
	\label{eq-parkpdf}
\end{equation}
Equation~(\ref{eq-parkpdf}) is not valid for $y<1.5$, but as we are
only considering the tail of the distribution, it suffices for our
purposes.  We show the tails of the functions $f\, P(y)$ for a 
Gaussian and texture PDF in Figure~\ref{fig:nongaus_pdfs}.  Note
that the texture PDF has a significantly wider tail for $y\gsim 2$.

\subsection{The Radius-Temperature Relation}

We next consider the relationship between the initial comoving radius
of the perturbation and the present temperature of a cluster.
We make the standard assumption that a cluster forms by undergoing
a spherical collapse to a singular isothermal sphere with a virial
radius equal to half its maximum expansion radius (e.g., White \etal\
1993, Eke \etal\ 1996, Pen 1997).  The X-ray temperature simply 
reflects the depth of the gravitational potential well of the
cluster.  We assume a hydrogen mass fraction of 0.76.  For clusters
collapsing at redshift $z_f$, we obtain
\begin{eqnarray}
	T & = & \frac{\mu m_p GM}{2r_{\rm vir}} \\
	  & = & 7.8 \mbox{\, keV} 	
		\left(\frac{M}{10^{15} h^{-1}
		M_\odot}\right)^{\twothirds}
		\left(\frac{\Delta_c(z_f)}{18\pi^2}\right)^{\onethird}
		\left(\frac{\Omega_0}{\Omega(z_f)}\right)^{\onethird}
		(1+z_f)
	\label{eq-ktmass}
	\\
	& = & 8.6 \mbox{\, keV\ } \Omega_0^{\twothirds}
		\left(\frac{R}{10\hmpc}\right)^2
		\left(\frac{\Delta_c(0)}{18\pi^2}\right)^{\onethird}
		\times
		\left[
		\left(\frac{\Delta_c(z_f)}{\Delta_c(0)}\cdot
			\frac{\Omega_0}{\Omega(z_f)}\right)^{\onethird}
		(1+z_f)
		\right]
	\label{eq-ktradfull}\\
	& \equiv & T_0(R) \times \phi\left(z_f\right),
	\label{eq-ktrad}
\end{eqnarray}
where the quantity in the square brackets is designated $\phi$,
and only depends on the background cosmology and the formation
redshift $z_f$.  The function $T_0(R)$ is the Radius-Temperature 
relation for collapse at $z_f=0$, for a given background cosmology,
thus $\phi(z_f=0) = 1$.  Equation~(\ref{eq-ktmass}) was derived
previously by Eke \etal\ (1996).
The factor $\Delta_c\equiv 3M/(4\pi r_{\rm vir}^3\rho_c)$ 
is the ratio of the virialized physical mass density
to the \emph{critical} (not the mean) 
cosmological density at the time of collapse,
and $r_{\rm vir}$ is taken to be $r_{\rm max}/2$.  In an $\Omega=1$,
matter dominated universe, $\Delta_c = 18\pi^2 \approx 178$; for $\Omega<1$, 
$\Delta_c$ can be derived from uniform spherical collapse (see 
Eke \etal\ 1996).  The quantity $\Delta_c$ depends only on 
the background cosmology.  Because $\Delta_c$ and $\Omega$ asymptote
to $18\pi^2$ and $1$ respectively as $z_f \rightarrow \infty$, 
and hence change by a factor $<\Omega_0^{-1}$,
the $z_f$-dependence of $\phi$ is dominated by the factor $1+z_f$.

Evrard \etal\ (1996) have compared numerical simulations for Gaussian
$\Omega_0=1.0$ and $\Omega_0=0.2$ models, and found
that $T$ scales as $M^{2/3}$, as equation~(\ref{eq-ktmass}) implies.
Fitting a power law to their results, and converting mass
to comoving scale implies $kT = (8.8\pm 0.8\mbox{\ keV})
(\Delta_c/18\pi^2)^{0.164} \Omega_0^{2/3}(R/8\hmpc)^2$.   
This equation is consistent with 
equation (\ref{eq-ktradfull}) for all three background 
cosmologies we consider, using the results 
of the next section to determine $z_f$ and $\phi(z_f)$.
The detailed simulations necessary 
to check the cluster temperatures for the texture scenario 
have not been done.
Bartlett (1997) notes, however, that for any given 
density profile of a spherical collapse, the power law
$T\propto M^{2/3} \propto R^2$ should be the same --- 
the only difference might be in the constant of proportionality
(e.g. through a different $\phi$-dependence).

Taking the derivative of equation~(\ref{eq-ktrad}) with respect to $R$
gives 
\begin{equation}
	\frac{dT}{dR} = T_0(R) \frac{d\phi}{dR} +
		\phi[z_f] \frac{dT_0}{dR} = 
        \frac{T}{R}\left(2 + \frac{d\ln\phi}{d\ln R}\right),
	\label{eq-dktdr}
\end{equation}
where the redshift of formation $z_f$ 
must be calculated as a function
of $R$.  Finally, inverting equation~(\ref{eq-ktrad}) gives 
$R$ as a function of the temperature:
\begin{eqnarray}
	R & = & 10 \hmpc\mbox{\ } \Omega_0^{-\onethird} 
		\left(\frac{T}{8.6\,{\rm keV}}\right)^{\onehalf}
		\left(\frac{18\pi^2}{\Delta_c(0)}\right)^{1/6}
		\times
		\left[ \left(
			\frac{\Delta_c(0)}{\Delta_c(z_f)}\cdot
			\frac{\Omega(z_f)}{\Omega_0}
			\right)^{1/6}
			\frac{1}{(1+z_f)^{\onehalf}} \right] 
	\label{eq-r_ktfull}\\
	&\equiv & 
		R_0(T)
		\times
		\phi\left[z_f(y)\right]^{-\onehalf},
	\label{eq-r_kt}
\end{eqnarray}
where $\phi$ is defined as above, the factor in front of the square
brackets has been replaced by $R_0(T)$, the Radius-Temperature
relation at $z_f=0$.  Note that equation~(\ref{eq-r_kt}) so defined 
is only valid when $z_f$ is known.

\subsection{Formation Epoch of Clusters in PS}

We now have another parameter to consider, the
redshift of formation $z_f$.  The standard assumption is to 
assume that the cluster has ``just formed'' at the redshift $z_0$ 
at which we observe it, i.e., that $z_f=z_0$.  
In the $\Omega_0=1$ Gaussian
CDM scenario, this assumption is valid, since clusters form late and
accretion continues to the present
(e.g., Lacey \& Cole 1993). 
If $\Omega_0<1$, however, linear 
growth of perturbations freezes out
when the universe enters its phase of free expansion
at $1+z \sim 1/\Omega_0$ (see \cite{PPC}).  Therefore, 
a given structure must collapse at a higher redshift 
than in the $\Omega_0=1$ case in order to form.  
If there is a sufficiently positively-skewed non-Gaussian tail to 
the distribution function, as there is for textures, 
a larger fraction of the universe collapses 
non-linearly early on (see Bartlett \etal\ 1993
and Pen \etal\ 1994).  Furthermore, for textures and other
``causal-seed'' models, whose density perturbations are compensated,
there is a limit to the final mass of a collapsing structure, defined
by the size of the  causal horizon at the time the perturbation is seeded.  
Because the integrated overdensity is zero on larger scales, 
no further accretion can occur (Bartlett \etal\ 1993). 

We now derive the mean redshift of formation $\langle z_f\rangle$,
and use that redshift in equation~(\ref{eq-ktrad}) to determine the
typical temperature of clusters as a function of $R$.
The net differential number density 
of clusters which \emph{form} in redshift interval 
$(z,z+dz)$ follows from equation~(\ref{eq-nr}):
\begin{equation}
\frac{dn_R}{dz} dz = 
	\frac{3}{4\pi R^3} \frac{\alpha}{R} f
		\frac{d}{dz}\left\{y(z)P\left[y(z)\right]\right\}dz
	\label{eq-dndz}
\end{equation}
where 
\begin{eqnarray}
y(z)  \equiv \frac{\delta_c(z)}{\sigma_R(z)} & = &
	\frac{\delta_c(0)}{\sigma_R(0)} \cdot 
		\frac{\delta_c(z)}{\delta_c(0)D(z)}  \\
	& \equiv & y_0 \frac{\delta_c(z)}{\delta_c(0)D(z)}.
	\label{eq-yzdef}
\end{eqnarray}
Here $y_0$ refers to the value of $y(z)$ at $z = 0$,
$D(z)$ is the linear growth factor normalized to unity at
redshift zero, and $\delta_c(z)$ is the critical linear 
density of collapse at redshift $z$.  For clusters 
observed at redshift $z_0$, the mean redshift of formation $\langle
z_f \rangle_{z_0}$ is obtained by averaging the redshift $z$ 
weighted by equation~(\ref{eq-dndz}):
\begin{eqnarray}
	\langle z_f \rangle_{z_0} & = &
 		\frac{\int_{z_0}^\infty z\frac{dn_R}{dz} dz}
			{\int_{z_0}^\infty \frac{dn_R}{dz} dz}\\
	& = & z_0 + \frac{1}{y(z_0)P\left[y(z_0)\right]}
			\int_{z_0}^\infty y(z)P\left[y(z)\right]dz
	\label{eq-zfdef}
\end{eqnarray}
where the second line comes from integrating the numerator
by parts, integrating the denominator, and cancelling terms.
The mean redshift of 
formation of clusters at $z_0$ is a function only of the 
background cosmology through $\delta_c(z)$ and $D(z)$,
and the current value of $y_0$ through equation~(\ref{eq-yzdef}).  
We take the formation redshift required in equation~(\ref{eq-ktrad})
to be $\langle z_f\rangle$.  
When using equation~(\ref{eq-zfdef}) with 
the present epoch abundances below, we set $z_0=0$.

\subsection{Determining $\eta$ from X-ray Cluster Surveys}

Using equations~(\ref{eq-ktrad}) and (\ref{eq-dktdr}) 
in equation~(\ref{eq-ntps}) gives
\begin{equation}
	2 n_T T \left(1 + \frac{\alpha}{2}\frac{d\ln\phi}{d\ln y} \right)
	= \frac{3}{4\pi R^3}\alpha y f\, P(y)
\end{equation}
where $y$ is now understood to be its present day value, and
we have used $\frac{d\ln y}{d\ln R} = -\frac{d\ln \eta}{d\ln R} = \alpha $, 
because of our power-law assumption.  Replacing
$R$ with equation~(\ref{eq-r_kt}) and separating those quantities
depending on $T$ from those depending on $y$ gives
\begin{equation}
	n_T T\frac{8\pi R^3_0(T)}{3} = \frac{\phi^{\frac{3}{2}}}
			{1 + \frac{\alpha}{2}\frac{d\ln\phi}{d\ln y}}
			\alpha y f\, P(y).
	\label{eq-ntcomp}
\end{equation}
We need now only to relate observations of clusters to the 
left-hand-side of equation~(\ref{eq-ntcomp}).

\begin{figure}
\plotone{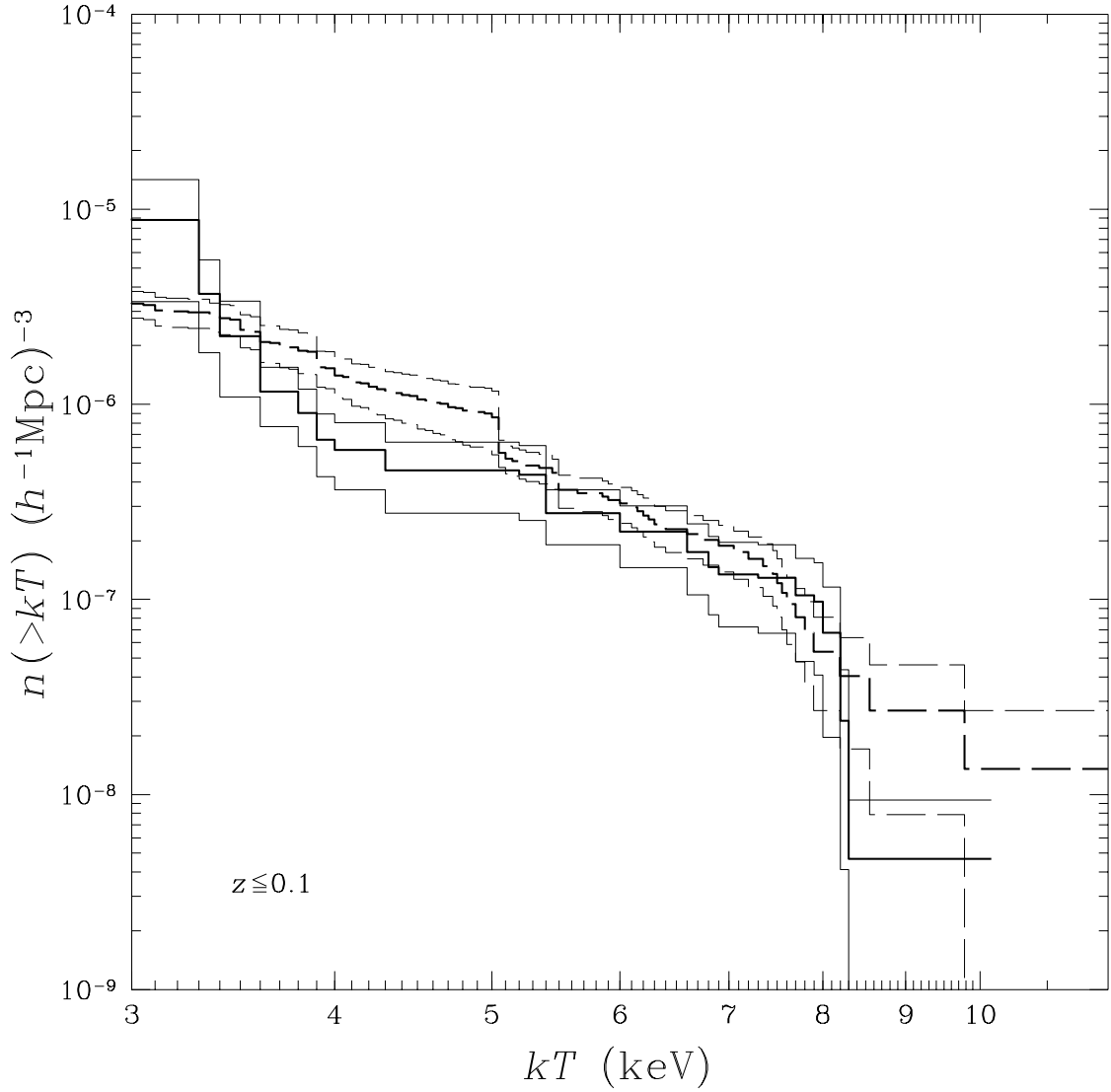}
\figcaption[nongaus_fig4.eps]{
Cumulative temperature function from Henry \& Arnaud (1991) 
(thick solid line), and 
from the XBACS survey (\cite{xbacs}) (thick dashed line), 
both with $1\sigma$ Poisson errors from equation~(\ref{eq-nktsigobs}) 
(thin histograms).  
\label{fig:nongaus_HA_xbac_comp}
}
\end{figure}

There are two virtually complete X-ray flux-limited surveys 
of rich galaxy clusters at low redshift.  
The first was compiled by Henry \& Arnaud (1991), 
and the second is the ``X-ray Brightest Abell Cluster Survey''
(XBACS) from the ROSAT all-sky survey (\cite{xbacs}).  The
Henry \& Arnaud (1991) survey is essentially contained 
within the XBACS survey.
An unbiased estimator of the number density of clusters in the 
temperature interval $(T,T+\Delta T)$ for a flux-limited survey is given by
\begin{equation}
	n_T \Delta T = \sum_{T < T_i < T+\Delta T }V^{-1}_{{\rm max},i},
	\label{eq-nktobs}
\end{equation}
where $V_{{\rm max},i}$ is the maximum volume in which the cluster of
temperature $T_i$ (in keV)
could be detected given the flux limit and geometric boundaries of the
survey.  In this case, we require a geometrical boundary at 
galactic latitude $|b|>20^\circ$, and redshift $z \leq 0.1$. 
The full XBACS sample takes the maximum redshift to 
be $z_{\rm max}=0.2$.  
The Poisson variance, following Pen (1997), is given by:
\begin{equation}
	\sigma^2\left(n_T\Delta T\right) 
		=  \sum_{T < T_i < T+\Delta T }
		V^{-2}_{{\rm max},i}.
	\label{eq-nktsigobs}
\end{equation}
The cumulative temperature functions $n(> T)$ for the two samples are
shown in Figure~\ref{fig:nongaus_HA_xbac_comp}.
The two temperature functions are
quite consistent in the overlapping temperature range
$>3\,$keV, so we use this part of the 
($10\times$) larger XBACS survey for our analysis.  

Substituting equation~(\ref{eq-nktobs}) gives, finally, 
\begin{equation}
	\left(\frac{T}{\Delta T}
	\sum_{T < T_i < T+\Delta T }V^{-1}_{{\rm max},i}\right)
 \frac{8\pi R^3_0(T)}{3} = \frac{\phi^{\frac{3}{2}}\alpha y f\, P(y)}
			{1 + \frac{\alpha}{2}\frac{d\ln\phi}{d\ln y}}.
	\label{eq-yfromobs}
\end{equation}
We define the temperature bins ($T,\ T+\Delta T$) 
so that there are at least two clusters in each.
For a fixed background cosmology and power law index $\alpha$, 
the left-hand side of equation~(\ref{eq-yfromobs}) contains only observed 
quantities, and the right-hand side is a function of $y$ and the PDF.  
For each PDF, we therefore can solve for $y$ implicitly 
as a function of $T$.
We then convert $T$ to $R$ using equation~(\ref{eq-r_kt}) and our
derived value of $y$, and obtain a value of
$\sigma_R^{\rm clusters} \equiv \delta_c/y$ as a function of the value
of $R$ inferred
from each temperature bin.  
Multiplying by $\Omega_0^{0.6}$ gives the cluster-inferred $\e{R}^{\rm
clusters}$, which can be compared directly with the values from the
rms measurements.  

A clear comparison with rms-inferred $\eta$ can be made if we
extrapolate all values of $\eta$ to a common scale, using the power
law model we have assumed throughout.  We choose $10\hmpc$ because
this scale corresponds to a temperature within the range of the
observations for $\Omega_0=0.3-1.0$: 8.6 keV for $\Omega_0=1$ and
$3.4\sim 4.0$ keV (depending on $\phi$) for $\Omega_0=0.3$.  For
$\Omega_0 = 0.3$, the scale $R=8\hmpc$ corresponds to a temperature of
$2.2\sim 2.6$ keV, outside the range of observations we consider.

For each temperature bin, we therefore extrapolate the derived 
value of $\e{R}$ to $10\hmpc$ using 
$\e{R} \propto \sigma_R \propto R^{-\alpha}$, 
and obtain a measurement of $\e{10}$ for each bin.  
We wish to find the resulting probability distribution in $\e{10}$.  
We find that the statistical 
errors in $\e{10}$ from equations~(\ref{eq-nktsigobs}) and
(\ref{eq-yfromobs}) are smaller than the scatter of the values of
$\e{10}$, though the discrepancies are not as drastic as in the case
of the rms measurements.  Certainly, then, 
there remains some type of systematic error.
We sum the distribution functions of each extrapolated $\e{10}$ to
obtain the total distribution function, assuming each error is
Gaussian:
\begin{equation}
	{\cal P}(\e{10}|D_{\rm clus}~\mbox{PDF}) =
	\frac{1}{N} \sum_{i=1}^{N} 
		\frac{1}{\sqrt{2\pi\sigma_i^2}} 
		e^{-\frac{1}{2}\left(
			\frac{\e{10,i}-\e{10}}{\sigma_i}\right)^2},
	\label{eq-probclus_def}
\end{equation}
where the sum is over the temperature bins labeled by $i$.
Note that we are summing over the individual distributions because
the individual temperature bins are not independent measures of
$\e{10}$; i.e., the data set as a whole is taken as a single
measurement, with the error given by the total distribution
${\cal P}(\e{10}|D_{\rm clus}~\mbox{PDF})$.
  
\begin{figure}
\plotone{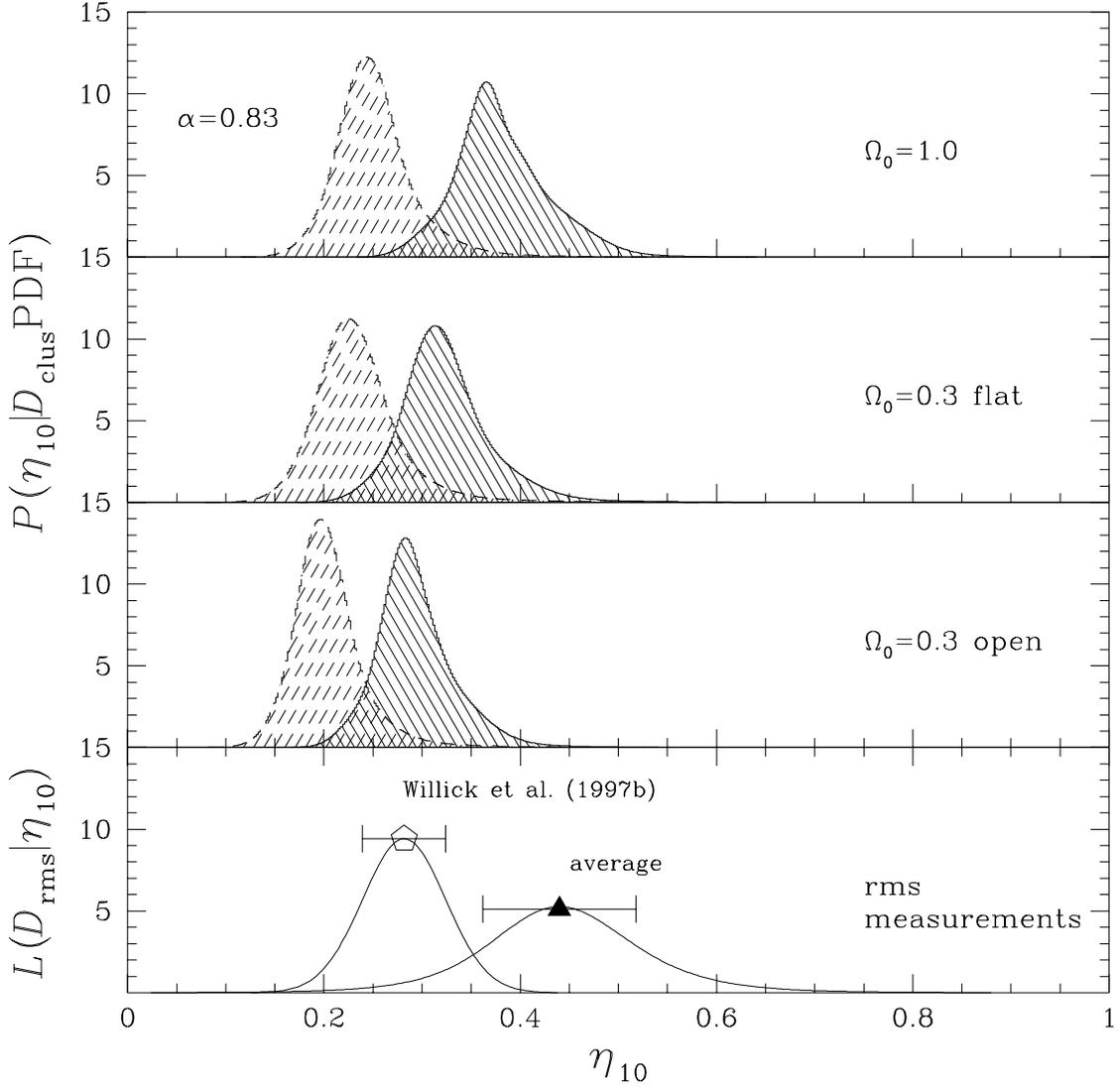}
\figcaption[nongaus_fig5.eps]{
Cluster-inferred $\e{10}$ compared with peculiar velocity/redshift
distortion-inferred $\e{10}$ at 10$\hmpc$ for $\alpha=0.83$, for
various assumed background cosmologies.
The solid-filled distribution functions are for a Gaussian PDF, and
the dashed-filled distribution functions are for a texture PDF.  
The filled triangle is the maximum likelihood 
average of the measurements in Table~\ref{tab:rms-eta}.  
The open pentagon is the Willick \etal\ (1997b) result.
\label{fig:nongaus_0.83_newhist}
}
\end{figure}

\begin{figure}
\plotone{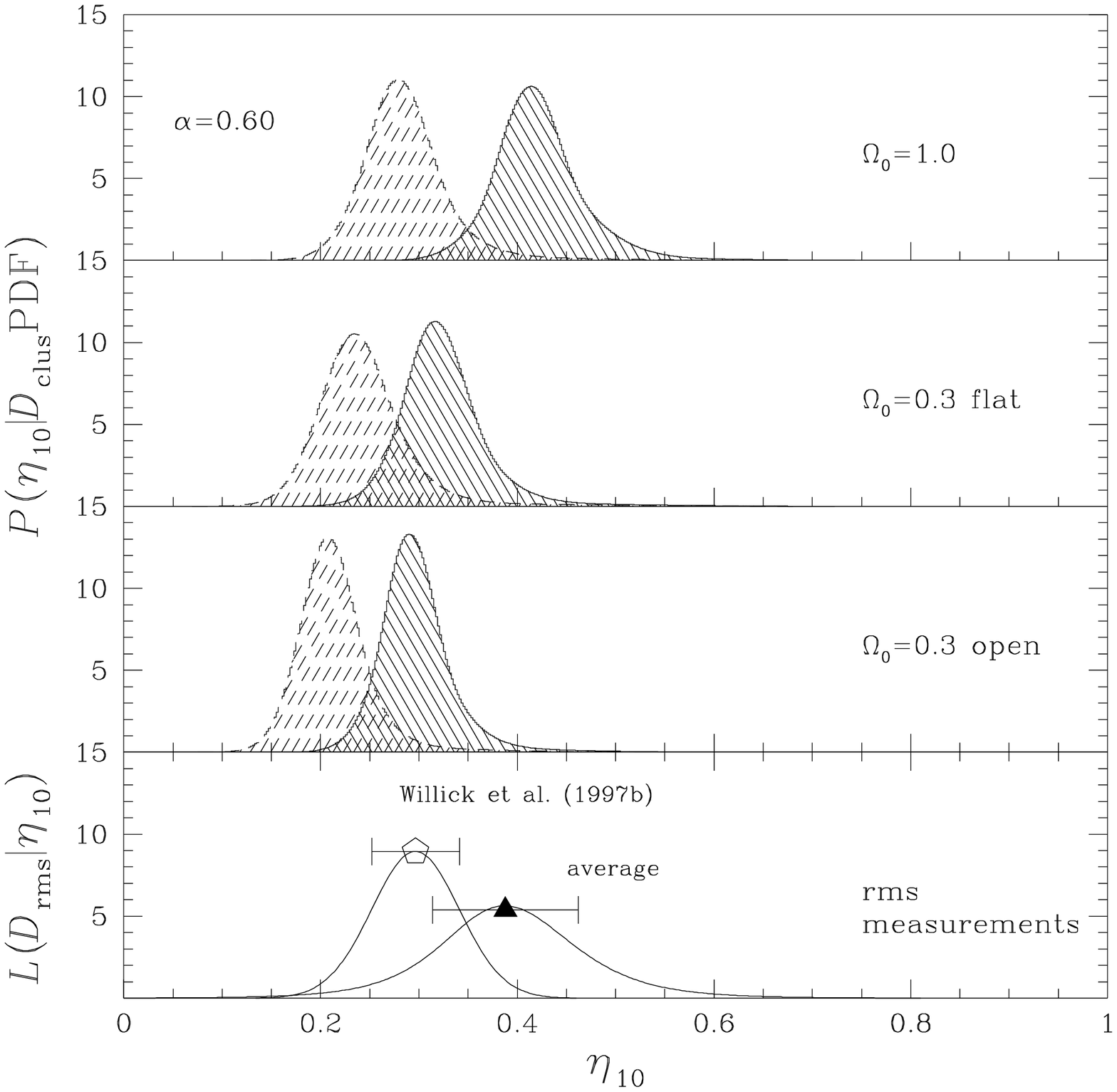}
\figcaption[nongaus_fig6.eps]{
Cluster-inferred $\e{10}$ compared with peculiar velocity/redshift
distortion-inferred $\e{10}$ at 10$\hmpc$ for $\alpha=0.60$.
The solid-filled distribution functions are for a Gaussian PDF, and
the dashed-filled distribution functions are for a texture PDF.  
\label{fig:nongaus_0.60_newhist}
}
\end{figure}

\begin{figure}
\plotone{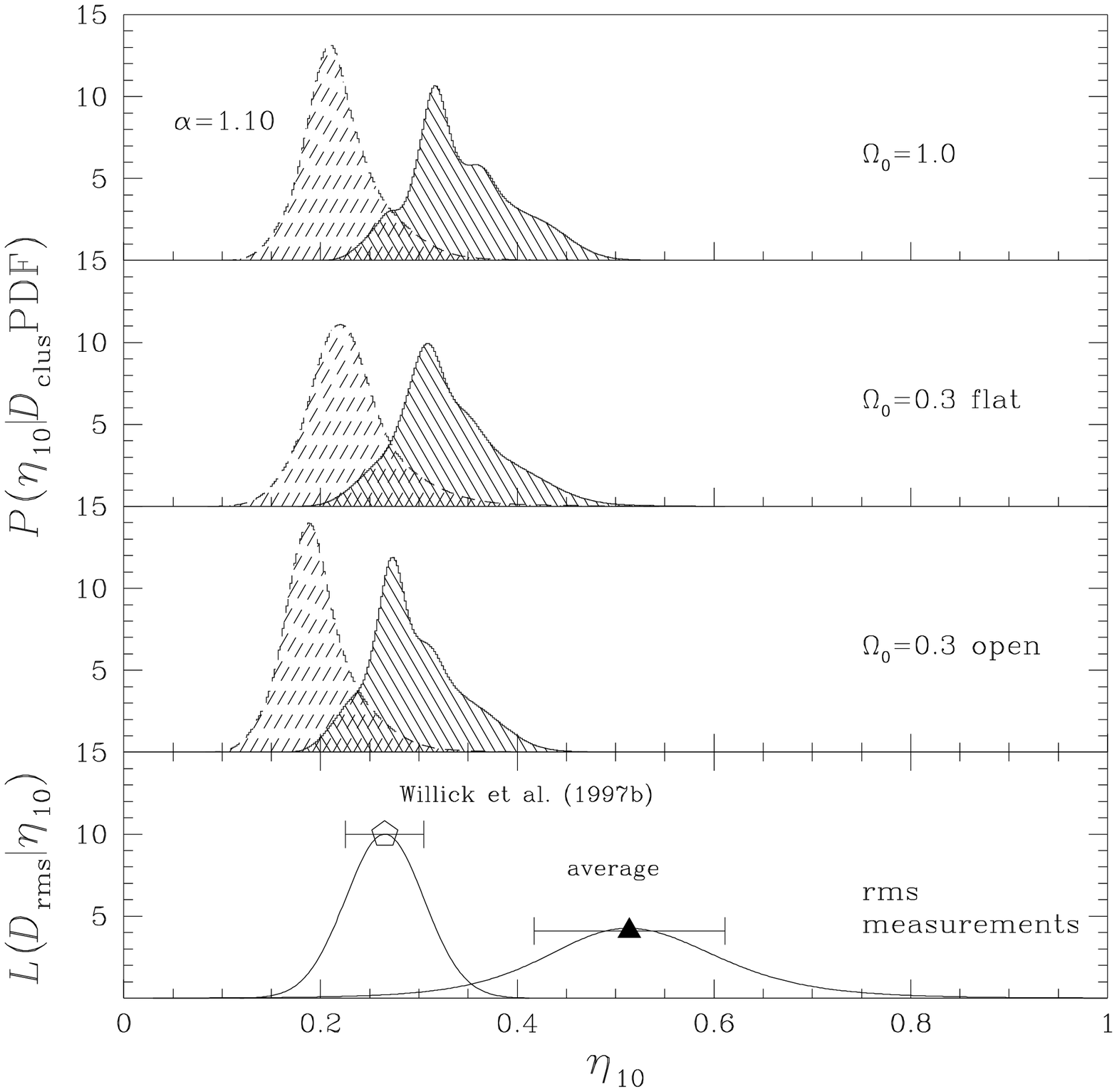}
\figcaption[nongaus_fig7.eps]{
Cluster-inferred $\e{10}$ compared with peculiar velocity/redshift
distortion-inferred $\e{10}$ at 10$\hmpc$ for $\alpha=1.10$.
The solid-filled distribution functions are for a Gaussian PDF, and
the dashed-filled distribution functions are for a texture PDF.  
\label{fig:nongaus_1.10_newhist}
}
\end{figure}

Figures~\ref{fig:nongaus_0.83_newhist}-\ref{fig:nongaus_1.10_newhist}
show the cluster-inferred 
$\e{10}$ distributions for power 
laws $\alpha=0.83,$\ 0.60, and 1.10, and for
different values of $\Omega_0$ and the PDF.  
Also shown are the likelihoods for mean peculiar velocity/redshift 
distortion-derived rms values discussed in \S 2, 
and the Willick \etal\ (1997b) result alone.
The means and standard deviations of the resulting distribution of
$\e{10}$ measurements are given in Table~\ref{tab:cluster-eta}, 
for various values of $\Omega_0$ and $\alpha$. 

The most prominent feature of the inferred rms from clusters is its
strong dependence on the PDF{}.  Gaussian and texture models give
clearly different predictions for any given cluster data set.  The
degree to which the cluster-inferred value of $\e{10}$ differs between
Gaussian and texture PDFs is directly related to the degree to which
the tail of the texture PDF is distinguishable from a Gaussian.

\section{Discussion}

\subsection{Which PDF is More Likely?}

	We quantify the comparison between the rms and
cluster-inferred determinations of $\e{10}$ with a 
Bayesian model comparison analysis (Loredo 1990).  
For now, we fix the background cosmology
and the power law index $\alpha$, and suppress the 
dependence on these assumptions.
We ultimately seek the probability that a certain PDF is 
correct, given the cluster data $D_{\rm clus}$ and the
rms data $D_{\rm rms}$:
\begin{equation}
{\cal P}(\mbox{PDF}|D_{\rm clus}~D_{\rm rms}) = 
	\frac{{\cal P}(\mbox{PDF})}
		{{\cal P}(D_{\rm clus}~D_{\rm rms})}
	{\cal P}(D_{\rm clus}~D_{\rm rms}|\mbox{PDF}),
	\label{eq-probpdf_def}
\end{equation}
where we have used Bayes Theorem.  The fraction in front consists of
a prior (which we take as uniform) and a normalization factor.
Using the product rule, we separate the last probability:
\begin{equation}
	{\cal P}(D_{\rm clus}~D_{\rm rms}|\mbox{PDF})=
		{\cal P}(D_{\rm clus}|\mbox{PDF})
	{\cal P}(D_{\rm rms}|D_{\rm clus}~\mbox{PDF}).
	\label{eq-probjointdata}
\end{equation}
Since we are primarily interested in the rms-versus-tail 
comparison, we take the first probability in 
equation~(\ref{eq-probjointdata}) to be independent of the PDF.  A more
sophisticated analysis would evaluate the ``goodness of fit'' 
of the cluster data for each PDF, and thereby estimate the
absolute probability $	{\cal P}(D_{\rm clus}|\mbox{PDF})$.  
However, as mentioned in \S 3, the scatter in the inferred values of 
$\e{10,\rm clus}$ is greater than the calculated Poisson errors, 
so any such analysis will have to take into account unknown 
systematic errors (e.g. using the model described in \S 2.2).  

The dependence of $D_{\rm rms}$ on $D_{\rm clus}$ in the
second probability in equation~(\ref{eq-probjointdata}) 
is through the value of $\e{10}$; using 
the summation rule gives
\begin{equation}
	{\cal P}(D_{\rm rms}|D_{\rm clus}~\mbox{PDF})=
		\int d\e{10} {\cal P}(D_{\rm rms}|\e{10}~\mbox{PDF})
			{\cal P}(\e{10}|D_{\rm clus}~\mbox{PDF}).
	\label{eq-pintegral}
\end{equation}
The resulting expression for equation~(\ref{eq-probpdf_def}), 
suppressing all factors
assumed to be independent of the PDF, is
\begin{equation}
	{\cal P}(\mbox{PDF}|D_{\rm clus}~D_{\rm rms})  \propto
		\int d\e{10} {\cal P}(D_{\rm rms}|\e{10}~\mbox{PDF}) 
			{\cal P}(\e{10}|D_{\rm clus}~\mbox{PDF}).
\end{equation}

The cluster distribution function in $\e{10}$,
${\cal P}(\e{10}|D_{\rm clus}~\mbox{PDF})$, is given by 
equation~(\ref{eq-probclus_def}).  
The rms probability depends only the power law
slope $\alpha$, and the weighting of the different analyses, 
and is independent of the PDF.
If we use the error model in \S 2.2, the probability
density ${\cal P}(D_{\rm rms}|\e{10}~\mbox{PDF})$
is given by the likelihood function 
from equation~(\ref{eq-prob_eta}) and Figure~\ref{fig:nongaus_rmslike}, 
\begin{equation}
	{\cal P}(D_{\rm rms}|\e{10}~\mbox{PDF}) =
		{\cal L}_{\rm rms}(\e{10}).
	\label{eq-probrms}
\end{equation}
If we are considering only the Willick \etal\ (1997b) data, 
$D_{\rm Willick}$, and
we assume normal errors, we use
\begin{equation}
	{\cal P}(D_{\rm Willick}|\e{10}~\mbox{PDF}) =
		\frac{1}{\sqrt{2\pi\epsilon^2_{\rm W}}}
		\exp\left[-\frac{1}{2}
			\left(\frac{\e{10}-\e{\rm W}}
				{\epsilon_{\rm W}}\right)^2\right],
	\label{eq-probwillick}
\end{equation}
where $\e{\rm W}$ and $\epsilon_{\rm W}$ are given by   
the values inferred from their analysis, also listed in 
Table~\ref{tab:rms-eta}.

	Since we do not know the full range of possible PDFs, 
we take the Bayesian approach of considering the 
likelihood ratio (the ``odds'', or ``Bayes factor'' for uniform priors,
Loredo~1990) of a Gaussian versus a texture PDF.
The values of the likelihood ratio 
${\cal P}(\mbox{Gaussian})/{\cal P}(\mbox{texture})$ 
are listed in Table~\ref{tab:likelihood-ratio}, for 
various background cosmologies and values of $\alpha$.
If we take the average of the rms measurements, a Gaussian PDF is
favored for any background cosmology.  The probability of a Gaussian
PDF is always $>2\times$ that of a texture PDF.  The confidence level
of the accepting a Gaussian hypothesis is $70-88\%$, roughly
$1\sigma\sim 1.5\sigma$.  
If we only use the Willick \etal\ (1997b) measurement,
however, then which PDF is favored \emph{depends strongly} on the
background cosmology, and somewhat on the power law index $\alpha$.  
In the case of $\alpha=0.83$, if we live in an open $\Omega_0=0.3$ universe,
then the Gaussian PDF is $3\times$ more likely than is the texture
PDF; if we have a flat $\Omega_0=0.3$ universe ($\Lambda=0.7$), then a
Gaussian and texture PDF are roughly equally likely;
and if $\Omega_0=1$, 
a texture PDF is $3\times$ more likely than a Gaussian.  

\subsection{Constraints from Cluster Evolution?}

We have only considered the present day abundances of
clusters, and we have a degeneracy between the assumed PDF and the
background cosmology. 
The evolution of cluster abundances, however,
offers an opportunity to break this degeneracy.  In particular, high
$\Omega_0$ universes 
evolve more rapidly than those with low $\Omega_0$.
Gaussian models with $\Omega_0=1$ are already heavily 
disfavored because they predict too rapid an evolution
in cluster abundances out to $z\geq 0.3$ (Henry 1997; 
Donahue \etal\ 1997).  The fact that the Willick \etal\ (1997b)
measurement leads to the $\Omega_0=1$ Gaussian scenario 
being disfavored relative to a texture scenario simply reinforces the
case against a flat matter-dominated universe for Gaussian fluctuations.  

\begin{figure}
\plotone{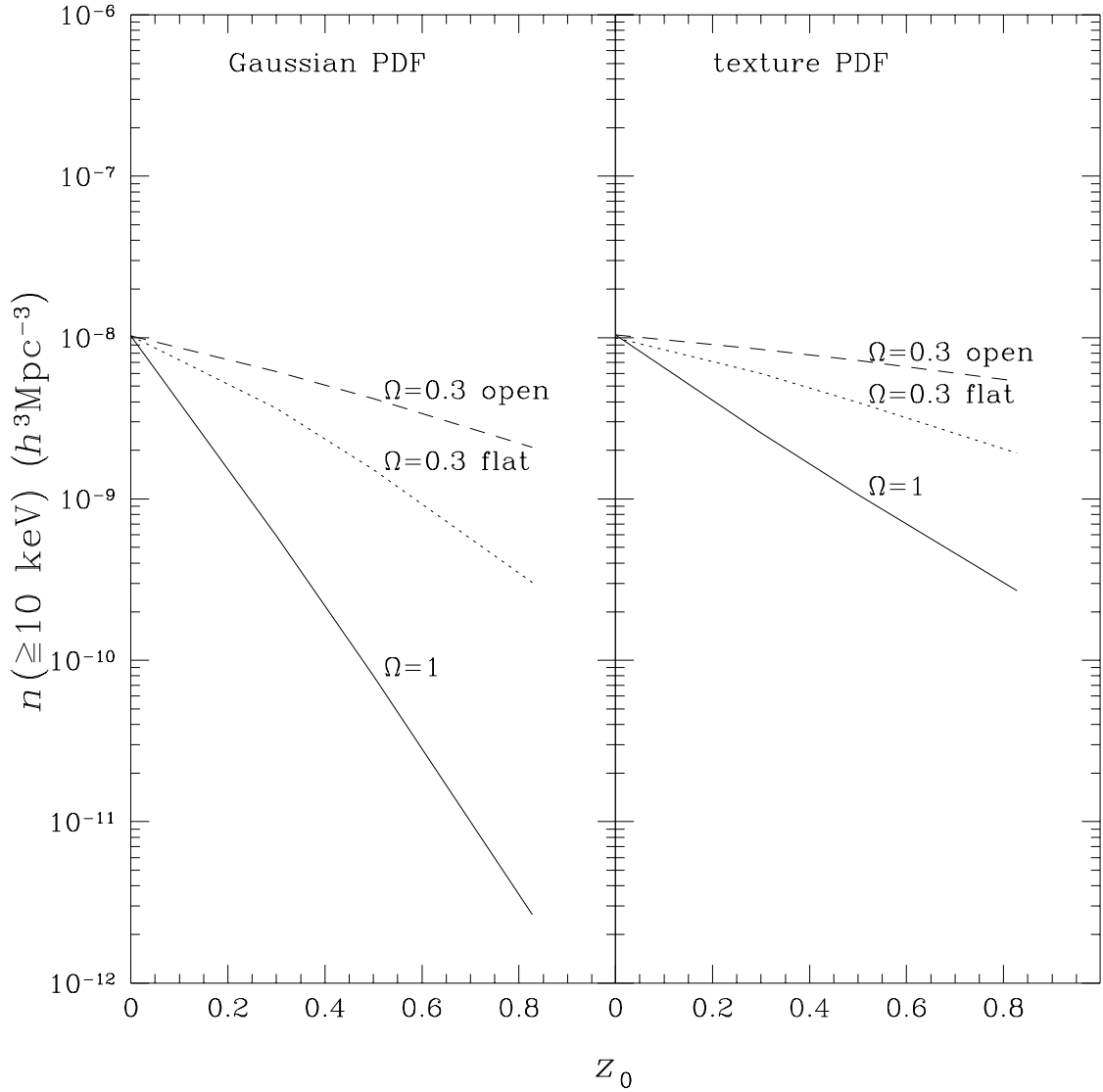}
\figcaption[nongaus_fig8.eps]{
Evolution of the number density of clusters with temperature
$\geq~10$~keV, normalized to the present day abundance, for Gaussian
and texture PDFs, and various background cosmologies.  
\label{fig:nongaus_10kevclusterevol}
}
\end{figure}

The formalism of \S 3 relating cluster abundances to the PDF
and the value of $\eta$ is fully generalizable to $z>0$.  
As an example, we show in Figure~\ref{fig:nongaus_10kevclusterevol} 
the cumulative number density of clusters 
$\geq 10$~keV, as a function of redshift for various
scenarios.  These curves are normalized to a present-day number density of
$1\times 10^{-8}\ h^3{\rm Mpc}^{-3}$ 
(see Figure~\ref{fig:nongaus_HA_xbac_comp}).  
For fixed background cosmology, 
the texture scenarios evolve slower than the Gaussian ones 
because of the shallower texture PDF.  For fixed PDF, the evolution of
the number density evolves fastest with $\Omega_0=1$ and slowest with 
open $\Omega_0=0.3$.  A Gaussian PDF with a
flat $\Omega_0=0.3$ background cosmology has roughly the same
evolutionary history as a texture PDF with $\Omega_0=1$.  
If evolution is observed to be significantly slower or more
rapid than this, then an $\Omega_0=1$ texture scenario can be ruled
out.   

Henry~(1997) reports that the number density of $z = 0.3$ clusters 
at the hot end of the temperature function (\gsim 6.5 keV) 
is a factor of $\sim 2.5$ lower than today (his Figure 2), though
this is still within the calculated scatter due to Poisson statistics.  
For $\Omega_0=1$, the Gaussian model predicts a factor
$\sim 8$ decrease in the density of such clusters at $z = 0.3$
while the texture model predicts a factor $\sim 3$ decrease.  For the 
$\Omega_0=0.3$ scenarios, the reduction factor is $1.1\sim 2$.
Henry~(1997) reports that the cluster evolutionary data 
rule out the Gaussian $\Omega_0 = 1$ at 99\% confidence.

Donahue \etal\ (1997) report an extremely hot X-ray temperature of 
$14.7\pm 4$ keV for cluster MS1054-03, at redshift 0.828. 
They indicate that this result, along with data from 
other high-redshift clusters they have studied, 
implies that, within errors of $\sim 50\%$, 
the number density of clusters with $T>10$ keV is 
\emph{essentially unchanged} from $z=0$ to $z=0.8$.  
If this result is confirmed, even a
texture PDF would be ruled out for $\Omega_0=1$, since it 
predicts a decrease in the number density by a factor of $\sim 30$.
Further observations at redshift $z=0.5\sim 1.0$ will 
almost certainly resolve the ambiguity between models with different 
$\Omega_0$.  

\section{Summary and Conclusions}

We show that comparing the amplitude of tail fluctuations indicated by
the abundances of clusters of galaxies, to the amplitude of 
the rms mass fluctuation as indicated by galaxy peculiar velocities
and redshift space distortions, can be a strong test of the
Gaussianity of the PDF of the initial density field, with
a weak dependence on the slope of the density fluctuation 
spectrum.

Observations of peculiar velocities and redshift distortions in the
linear regime can
measure the rms level of gravitational potential fluctuations present in
the universe in the combination $\e{R}\equiv\sigma_R\Omega_0^{0.6}$,
independent of galaxy bias.  
Various existing measurements of this quantity
are not consistent
with one another within their stated errors.
We assume the existence of a systematic error in each
measurement, and use a maximum likelihood technique to 
combine these recent measurements of $\e{R}$.  The systematic error
leads to a $\sim 20\%$ uncertainty in the value of $\e{R}$.
This procedure gives
$\e{10,{\rm rms}}=0.39\pm 0.07 - 0.51 \pm 0.10$, depending on
the assumed slope of the mass fluctuation spectrum.
We use this value, and one recent measurement by Willick \etal\ (1997b),
which indicates $\e{10}= 0.27-0.30\pm 0.04$, to compare with the
cluster-inferred measures of $\e{10}$.
We chose the Willick \etal\ (1997b) measurement 
because it is the most recent
analysis which explicitly filters close to the $10\hmpc$ scale on
which clusters form. 

The value of $\eta_{10}$ inferred from cluster abundances is calculated
using the Press-Schechter approach for both a Gaussian and a texture
PDF.  We find a simple relation for the typical redshift 
of formation of clusters in both models.  This is particularly
important in the texture scenario, where the
shallower tail of the PDF leads to significant early formation of
clusters.  This typical redshift is important in the relation between
the virial temperature of the clusters and the comoving radius of the
initial perturbation. 
For a texture PDF, the observed abundances of X-ray clusters implies
$\e{10}=0.20\pm 0.03 - 0.25\pm 0.04$ for $\Omega_0=0.3$, and
$\e{10}=0.22-0.29 \pm 0.03$ for $\Omega_0=1.0$.
Given a Gaussian PDF, on the other hand, 
the cluster abundances indicate $\e{10}= 0.29\pm 0.04 - 0.33 \pm 0.05$ for
$\Omega_0=0.3$ and $\e{10}=0.34\pm 0.05 - 0.42\pm 0.04$ for
$\Omega_0=1$.

We calculate the relative likelihoods of each PDF for a 
range of cosmological parameters. 
Using the maximum likelihood average of the rms measurements with
the cluster abundances implies the Gaussian model is favored for any
$\Omega_0$.  However, using the Willick \etal\ (1997b) 
rms measurement with the clusters implies that
a Gaussian is favored for open $\Omega_0=0.3$,
a Gaussian and a texture PDF of roughly equal likelihood for 
flat $\Omega_0=0.3$, and a texture PDF is favored for $\Omega_0=1$.
In each case, the accepted hypothesis is accepted at a 
$\sim 1.5\sigma$ level, with likelihood ratio $\gsim 3$.  
Preliminary constraints from cluster evolution indicate that
low $\Omega_0$ models are probably more viable than 
$\Omega_0$=1 models.  Our conclusions can are 
summarized in Table~\ref{tab:summary_table}, where a \chk~(X)
indicates the PDF is favored (disfavored) over the 
alternative by $\gsim 3\times$.  A $?$ next to the \chk or X 
indicates the factor is $\lsim 3$.

Our results imply that the textures+CDM scenario
faces some challenges from 
observations on $5-25h^{-1}$Mpc scales, \emph{completely independent} 
of CMB measurements.  
Although the current state of the measurements of the rms prevents us
from firmly distinguishing the Gaussian and texture models,
the principle behind this method is clear.  
The observed cluster abundance at $z = 0$ predicts
significantly different values for $\e{10}$ for the two models. 
More accurate direct
determinations of the measured rms from future redshift surveys and 
peculiar velocity measurements,
or from an improved understanding of the systematic effects which are
present in the current analyses, 
will surely enable us to \emph{definitively}
test \emph{any} distribution function of initial density perturbations.
Furthermore, upcoming high redshift X-ray cluster data 
will likely break any remaining degeneracy due 
to the assumed background cosmology.
Soon, perhaps, we will finally be able to distinguish different models
for the initial PDF of the universe at 10$\hmpc$.

\acknowledgements{We thank Avishai Dekel and Ed Turner for useful
conversations.  MAS acknowledges the support of the Alfred P. Sloan
Foundation, Research Corporation, and NSF Grant AST96-16901.}


\appendix
\section{What is the Correct Correspondence Between Tophat
and Gaussian Filtering?}

	One of the traditional ways of relating different filtering
schemes is by matching the volume over which they filter.  For tophat
and Gaussian, then, this would correspond to $4\pi R_T^3/3 = 
(2\pi R_G^2)^{3/2}$, or $R_T \approx R_G/1.6$.  While this might be
appropriate when relating the \emph{mass} of objects picked out by
these filter scales, it is not the best way of relating the 
wavelengths of fluctuations measured by each filter.
Here we derive a method of relating tophat- and
Gaussian-filtered fields by their maximum correlation.

	Consider a homogeneous, isotropic scalar field
$F({\mathbf{x}})$ and its Fourier transform in the following convention:
\begin{eqnarray}
	F({\mathbf{x}}) & = & \sum_{\mathbf{k}} \widetilde{F}({\mathbf{k}})
				e^{i\mathbf{k\cdot x}}
			= \frac{V}{(2\pi)^3}\int d^3k 
				\widetilde{F}({\mathbf k})
				e^{i\mathbf{k\cdot x}},\\
	\widetilde{F}({\mathbf{k}}) & = & \frac{1}{V}\int_V
				d^3x F({\mathbf{x}})
				e^{-i\mathbf{k\cdot x}}.
\end{eqnarray}
Let the power spectrum $P(k) \equiv \langle |\widetilde{F}({\mathbf{k}})|^2
\rangle$.  The (dimensionless) tophat and Gaussian filters 
and their Fourier transforms are then
\begin{eqnarray}
	W_{R_T}({\mathbf{x}}) & = & V\frac{3}{4\pi R_T^3} \quad 
		\mbox{if\ } |{\mathbf{x}}| < R_T, \quad\mbox{0 otherwise}, \\
 	\widetilde{W}_{R_T}({\mathbf{k}}) & = & \frac{3 j_1(k R_T)}{k R_T}
		\equiv \widetilde{W}_T(kR_T),
		\\
	W_{R_G}({\mathbf{x}}) & = & V \frac{1}{(2\pi R_G^2 )^{3/2}}
		\exp(-x^2/(2R_G^2)), \\
 	\widetilde{W}_{R_G}({\mathbf{k}}) & = & \exp(-k^2 R_G^2/2) 
		\equiv \widetilde{W}_G(kR_G),
\end{eqnarray}
where $j_1(x)$ is the spherical Bessel function of order 1.
The filtered fields are $F_{R_{T,G}}({\mathbf{x}})=V^{-1}\int 
d^3y F({\mathbf{x-y}}) W_{R_{T,G}}({\mathbf{y}})$.
Using the convolution theorem, the variance for tophat and 
Gaussian filters is:
\begin{equation}
	\sigma_{T,G}^2(R_{T,G}) = \frac{V}{(2\pi)^3}\int dk\ 4\pi k^2 P(k)
		\widetilde{W}^2_{T,G}(kR_{T,G}).
\end{equation}
How do we relate results with a tophat filter and those with a  
Gaussian filter?  We maximize the correlation between 
the two filtered fields, normalized by their respective standard
deviations:
\begin{equation}
	\rho(R_T,R_G) \equiv \left\langle
			\frac{F_{R_T}({\mathbf{x}})}{\sigma_T(R_T)}
			\frac{F_{R_G}({\mathbf{x}})}{\sigma_G(R_G)} 
		\right\rangle
		= \frac{1}{\sigma_T(R_T)\sigma_G(R_G)} 
			V\int \frac{dk\ 4\pi k^2}{(2\pi)^3}P(k)
			\widetilde{W}_T(kR_T)\widetilde{W}_G(kR_G).
\end{equation}
For Gaussian Random Fields, the one-point distribution of each of 
the filtered fields is Gaussian, and the joint distribution of 
the two fields is a bivariate Gaussian.  In this case 
the quantity $\rho(R_T,R_G)$ 
is \emph{precisely} the normalized correlation coefficient 
of the joint distribution of the two fields.  
The two filtering schemes, therefore, are most 
correlated when $\rho(R_T,R_G)$ is maximized.  
In Table~\ref{tab:filter-correlation}, we list the peak correlations for power-law power spectra
$P(k) \propto k^{n}$\ for $n=-2, -1, 0,\ \mbox{and}\ 1$.  For
decreasing power spectra ($n<0$), the correlation between tophat 
and Gaussian filtering is very good, greater than 87\%, with the
ratio $R_T/R_G = 1.75 \sim 2$.  

	It should be noted that we are not comparing the \emph{value}
of $\sigma$ between the two filtering schemes --- we are seeking the
scales at which the two filters are most
correlated.  For example, for a CDM, $\Omega_0=1,\ n=1,$\
power spectrum, using the maximum correlation, 12$\hmpc$\ Gaussian
smoothing corresponds to 22.7$\hmpc$\ tophat smoothing; but 
$\sigma_G(12)/\sigma_T(22.7) = 0.86$, while 
$\sigma_G(12)/\sigma_T(25.4) = 1.0$.  




\clearpage

\clearpage
\begin{deluxetable}{lccccc}
\tablecaption{
RMS Measurements of $\e{10}$
\label{tab:rms-eta}
}
\tablehead{
\colhead{Reference} & \colhead{R} & \colhead{$\e{R}$} 
	& \multicolumn{3}{c}{$\e{10}$} \\ 
\colhead{} & \colhead{} & \colhead{} 
	& \colhead{$\alpha=0.60$} & \colhead{$\mathbf\alpha= 0.83$} 
	& \colhead{$\alpha=1.10$} 
}
\startdata 
 \cite{Z96}	& 10 & $0.70\pm 0.09$ & $0.70\pm 0.09$ & $\mathbf 0.70\pm 0.09$
	& $0.70\pm 0.09$  \nl
 \cite{kd96}	& 23 & $0.32\pm 0.11 $ & $0.53\pm 0.18$  
	& $\mathbf 0.64\pm 0.22$ & $0.80\pm 0.27$ \nl
 \cite{W97b}	& 8 & $0.34\pm 0.05$ & $0.30\pm 0.04$ 
	& $\mathbf 0.28\pm 0.04$ & $0.27\pm 0.04$  \nl
 \cite{Sigad97}	& 23 & $0.25\pm 0.04 $ & $0.41\pm 0.07$ 
	& $\mathbf 0.50\pm 0.08$ & $0.62\pm 0.10$ \nl
 \cite{Fisher94b} & 12.5 & $0.21\pm 0.13$ & $0.25\pm 0.15$ 
	& $\mathbf 0.26\pm 0.16$ & $0.27\pm 0.17$ \nl
 Cole \etal\ 1995a,b & 35 & $0.11\pm 0.03$ & $0.22\pm 0.06$ 
	& $\mathbf 0.30\pm 0.08$ & $0.42\pm 0.12$ \nl
 \cite{FSL94} & 30 & $0.20\pm 0.03$ & $0.39\pm 0.06$ 
	& $\mathbf 0.50\pm 0.07$ & $0.67\pm 0.10$ \nl
 {\bf Maximum Likelihood Avg} & 10 & ---	& $0.39\pm 0.12$ 
	& $\mathbf 0.44\pm 0.12$ & $0.51\pm 0.15$ \nl 
\enddata
\end{deluxetable}

\begin{deluxetable}{lccc}
\tablecaption{
XBACS Cluster Measurements of $\e{10}$
\label{tab:cluster-eta}
}
\tablehead{
\colhead{Cluster Scenario} & \multicolumn{3}{c}{$\e{10}$} \\ 
\colhead{} & \colhead{$\alpha=0.60$} & \colhead{$\mathbf\alpha= 0.83$} 
	& \colhead{$\alpha=1.10$}
}
\startdata
 Gaussian $\Omega_0=1$ & $0.423\pm 0.036$ & $\mathbf 0.381\pm 0.041$
	& $0.342\pm 0.049$ \nl 
 texture $\Omega_0=1$ & $0.287\pm 0.033$ & $\mathbf 0.252\pm 0.031$
	& $0.220\pm 0.032$ \nl 
 Gaussian $\Omega_0=0.3$\ flat & $0.326\pm 0.035$ & $\mathbf 0.325\pm 0.038$
	& $0.329\pm 0.047$ \nl
 texture $\Omega_0=0.3$\ flat & $0.245\pm 0.037$ & $\mathbf 0.235\pm 0.034$
	& $0.231\pm 0.036$ \nl
 Gaussian $\Omega_0=0.3$\ open & $0.298\pm 0.029$ & $\mathbf 0.293\pm 0.033$
	& $0.293\pm 0.042$ \nl
 texture $\Omega_0=0.3$\ open & $0.215\pm 0.028$ & $\mathbf 0.204\pm 0.027$
	& $0.198\pm 0.029$ \nl
\enddata
\end{deluxetable}

\begin{deluxetable}{llccc}
\tablecaption{Probability of Gaussian (G) Relative to Texture (T) PDF
\label{tab:likelihood-ratio}
}
\tablehead{
 \colhead{rms} & \colhead{$\Omega_0$ and}  
	& \multicolumn{3}{c}{$\cal P$(G)/$\cal P$(T) for $\alpha=$} \\
 \colhead{Averaging} &  \colhead{Geometry} 
	& \colhead{0.60} & \colhead{\bf 0.83} & \colhead{1.10} 
}
\startdata
 Maximum  &  1.0 	& 2.28 & {\bf 6.52} & 7.00 \nl
 Likelihood & 0.3 flat 	& 3.28 & {\bf 4.18} & 4.41 \nl
	& 0.3 open 	& 3.98 & {\bf 4.71} & 4.62 \nl
Willick \etal\  
	& 1.0 		& 0.124 & {\bf 0.318} & 0.704 \nl
(1997b) only 
	& 0.3 flat 	& 1.53 & {\bf 1.25} & 0.789 \nl
 	& 0.3 open 	& 3.28 & {\bf 3.01} & 2.13 \nl
\enddata
\end{deluxetable}

\begin{deluxetable}{lcccccc}
\tablecaption{
Summary of Tests of Gaussian (G) Versus Textures (T)
\label{tab:summary_table}
}
\tablehead{
\colhead{Tests} & \multicolumn{2}{c}{$\Omega_0=1$} 
	& \multicolumn{2}{c}{$\Omega_0=0.3~{\rm flat}$} 
		& \multicolumn{2}{c}{$\Omega_0=0.3~{\rm open}$} \\
\colhead{} & \colhead{G} & \colhead{T} & \colhead{G} & \colhead{T} 
	& \colhead{G} & \colhead{T} 
}
\startdata
rms average and $z=0$ clusters & \chk & X & \chk & X 
	& \chk & X \nl
Willick \etal\ (1997b) and $z=0$ clusters 
	& X & \chk & ?= & ?= 
	& \chk  & X \nl
cluster evolution & X & ?X & ?\chk & \chk 
	& \chk & \chk \nl
\enddata
\tablecomments{A ``\chk''~(``X'') indicates the PDF 
is favored (disfavored) over the alternative by 
$\gsim 3\times$ or $\gsim 1.5\sigma$; a ``?='' indicates
roughly equal likelihood; a ``$?$''\ next to the ``\chk''\ (``X'') 
indicates a $\lsim 3$ or $\lsim 1.5\sigma$ result
in favor (against).}
\end{deluxetable}

\begin{deluxetable}{ccc}
\tablenum{A1}
\tablecaption{
Correspondence Between Tophat and Gaussian Filtering
\label{tab:filter-correlation}
}
\tablehead{
\colhead{index $n$} & \colhead{$R_T/R_G$} & \colhead{$\rho(R_T,R_G)$} \\
\colhead{$(P(k)\propto k^n)$}  & \colhead{(maximum correlation)} 
	& \colhead{} 
}
\startdata
 $-2$	& 1.94 & 0.99 \nl
 $-1$ 	& 1.85 & 0.97 \nl
  $0$	& 1.75 & 0.87 \nl
  $1$	& 1.66 & 0.57 \nl
\enddata
\end{deluxetable}

\end{document}